\documentclass[preprint,aps,superscriptaddress,showpacs,preprintnumbers,amsmath,
amssymb,floatfix,12pt,nofootinbib]{revtex4}

\usepackage{color} 

\usepackage{graphicx}
\newcommand{\bea}{\begin{eqnarray}}
\newcommand{\eea}{\end{eqnarray}}

\def\etal{{\it et al}.\ }

                                {\end{figure}}

                                {\end{table}}

 \def\etal{{\it et al.}}
 \def\bar{\overline}
 \def\spose#1{\hbox to 0pt{#1\hss}}
 \def\gtapprox{\mathrel{\spose{\lower 3pt\hbox{$\mathchar"218$}}
 \raise 2.0pt\hbox{$\mathchar"13E$}}}

\newcommand*{\CPT}{\raise0.4ex\hbox{$\chi$}PT}
\newcommand*{\chpt}{\raise0.4ex\hbox{$\chi$}PT}
\newcommand*{\schpt}{S\raise0.4ex\hbox{$\chi$}PT}

\def\eqref#1{{(\ref{#1})}}

\def\bar{\overline}
\def\hat{\widehat}

\def\etal{{\it et al.}}
\def\bea{\begin{eqnarray}}
\def\eea{\end{eqnarray}}

\begin{document}
\bibliographystyle{apsrev}


\title{The \boldmath$\overline{B} \rightarrow D^* \ell \overline{\nu}$ 
form factor at zero recoil from three-flavor lattice QCD:  A model independent determination of $|V_{cb}|$}

\author{C.~Bernard} 
\affiliation{Department of Physics, Washington University,
St.~Louis, Missouri, USA}
\author{C.~DeTar} 
\affiliation{Physics Department, University of Utah,
Salt Lake City, Utah, USA}
\author{M.~Di Pierro}
\affiliation{School of Computer Science, Telecommunications and Information
Systems, DePaul University, Chicago, Illinois, USA}
\author{A.~X.~El-Khadra}
\affiliation{Physics Department, University of Illinois,
Urbana, Illinois, USA}
\author{R.~T.~Evans}
\affiliation{Physics Department, University of Illinois,
Urbana, Illinois, USA}
\author{E.~D.~Freeland}
\affiliation{Liberal Arts Department, The School of the Art Institute
of Chicago, Chicago, Illinois, USA}
\author{E.~Gamiz}
\affiliation{Physics Department, University of Illinois,
Urbana, Illinois, USA}
\author{Steven Gottlieb} 
\affiliation{Department of Physics, Indiana University,
Bloomington, Indiana, USA}
\author{U.~M.~Heller}
\affiliation{American Physical Society, 
Ridge, New York, USA}
\author{J.~E.~Hetrick}
\affiliation{Physics Department, University of the Pacific,
Stockton, California, USA}
\author{A.~S.~Kronfeld}
\affiliation{Fermi National Accelerator Laboratory, Batavia, Illinois, USA}
\author{J.~Laiho} 
\affiliation{Department of Physics, Washington University,
St.~Louis, Missouri, USA}
\affiliation{Fermi National Accelerator Laboratory, Batavia, Illinois, USA}
\author{L.~Levkova}
\affiliation{Physics Department, University of Utah,
Salt Lake City, Utah, USA}
\author{P.~B.~Mackenzie} 
\affiliation{Fermi National Accelerator Laboratory, Batavia, Illinois, USA}
\author{M.~Okamoto}
\affiliation{Fermi National Accelerator Laboratory, Batavia, Illinois, USA}
\author{J.~Simone}
\affiliation{Fermi National Accelerator Laboratory, Batavia, Illinois, USA}
\author{R.~Sugar}
\affiliation{Department of Physics, University of California,
Santa Barbara, California, USA}
\author{D.~Toussaint}
\affiliation{Department of Physics, University of Arizona,
Tucson, Arizona, USA}
\author{R.~S.~Van~de~Water}
\affiliation{Fermi National Accelerator Laboratory, Batavia, Illinois, USA}
\collaboration{Fermilab Lattice and MILC Collaborations}
\noaffiliation

\begin{abstract}
We present the first lattice QCD calculation of the form factor for
$\overline{B}\rightarrow D^* \ell \overline{\nu}$ with 
three flavors of sea quarks.  We use an improved staggered action for the light valence and sea quarks (the MILC
configurations), and the Fermilab action for the heavy quarks.  The
form factor is computed at zero recoil using a new double ratio
method that yields the form factor more directly than the previous
Fermilab method.  Other improvements over the previous calculation include the use of much lighter light quark masses, and the use of lattice (staggered) chiral perturbation theory in order to control the light quark discretization errors and chiral extrapolation.  We obtain for the form factor, ${\cal F}_{B\to D^*}(1)=0.921(13)(20)$, where the first error is statistical and the second is the sum of all systematic errors in quadrature.   Applying a $0.7\%$ electromagnetic correction and taking the latest PDG average for ${\cal F}_{B\to D^*}(1)|V_{cb}|$ leads to $|V_{cb}|=(38.7 \pm 0.9_{exp} \pm 1.0_{theo})\times 10^{-3}$.
\end{abstract}

\date{\today}

\maketitle

\section{Introduction}

The Cabibbo-Kobayashi-Maskawa matrix element $V_{cb}$ plays an important
role in the study of flavor physics \cite{Barberio:2007cr}.  Since
$|V_{cb}|$ is one of the fundamental parameters of the Standard Model,
its value must be known precisely in order to search for new physics by
looking for inconsistencies between Standard Model predictions and
experimental measurements.  For example, the Standard Model 
contribution to
the kaon mixing parameter $\epsilon_K$ depends sensitively on $|V_{cb}|$ 
(as the fourth power), and the present
errors on this quantity contribute errors to the theoretical prediction
of $\epsilon_K$ that are
around the same size as the errors due to $B_K$, the kaon bag parameter,
 which has been the focus of much recent work
\cite{Gamiz:2006sq, Bae:2005su, Antonio:2007pb, Aubin:2007pt}. It
is possible to obtain $|V_{cb}|$ from both inclusive and exclusive
semileptonic $B$ decays, and both determinations are limited by
theoretical uncertainties. The inclusive method \cite{Chay:1990da, Bigi:1992su, Bigi:1992ne, Bigi:1993fe, Bigi:1997fj} makes use of the
heavy-quark expansion and perturbation
theory.  The method also requires non-perturbative input from
experiment, which is obtained from the measured moments of the inclusive
form factor $\overline{B}\to X_c\ell \overline{\nu}_{\ell}$ as a
function of the minimum electron momentum.
The dominant uncertainties in this method are the truncation of the
heavy quark expansion and perturbation theory 
\cite{Buchmuller:2005zv,Bauer:2004ve}.
In order to be competitive with the inclusive determination of
$|V_{cb}|$ and thus serve as a cross-check, the exclusive method
requires a reduction in the uncertainty of the $B\to D^*$ semileptonic
form factor ${\cal F}_{B\rightarrow D^*}$, which has been calculated
previously using lattice QCD in the quenched approximation
\cite{Hashimoto:2001nb}.

Given the phenomenological importance of $|V_{cb}|$, we have revisited
the calculation of ${\cal F}_{B\rightarrow D^*}$ at zero recoil using
the 2+1 flavor MILC ensembles with improved light
staggered quarks \cite{Bernard:2001av, Aubin:2004wf}.  The systematic error due to
quenching is thus eliminated.  The systematic error associated with the
chiral extrapolation to physical light quark masses is also reduced
significantly.  Since staggered quarks are computationally less
expensive than many other formulations, we are able to simulate at quite
small quark masses; our lightest corresponds to a pion mass of roughly
$240$ MeV. Given the previous experience of the MILC Collaboration with
chiral fits to light meson masses and decay constants
\cite{Aubin:2004fs}, we are
in a regime where we expect
rooted staggered chiral perturbation theory (rS$\chi$PT)
\cite{Lee:1999zx, Aubin:2003mg, Aubin:2003uc, Sharpe:2004is,
Aubin:2004xd} to apply.  We therefore use the rS$\chi$PT result for the
$B\to D^*$ form factor \cite{Laiho:2005ue} to perform the chiral
extrapolation and to remove discretization effects particular to staggered quarks.  In addition, we
introduce a set of ratios that allows us to disentangle light- and
heavy-quark discretization effects, and we suggest a strategy for future
improvement.  Finally, we extract the $B\to D^*$ form factor using a
different method from that originally proposed in
Ref.~\cite{Hashimoto:2001nb}.  This new method requires 
many
fewer three-point correlation functions, and has allowed for a savings
of roughly a factor of ten in computing resources, while at the same
time simplifying the analysis.

The differential rate for the semileptonic decay $\overline{B}\to
D^*\ell\overline{\nu}_\ell$ is
\bea
\frac{d\Gamma}{dw} &=&
\frac{G^2_F}{4\pi^3}m^3_{D^*}(m_B-m_{D^*})^2\sqrt{w^2-1} \ {\cal
G}(w)|V_{cb}|^2|{\cal F}_{B\rightarrow D^*}(w)|^2,
\eea
where $w=v' \cdot v$ is the velocity transfer from the
initial state to the final state, and ${\cal G}(w)|{\cal F}_{B
\rightarrow D^*}(w)|^2$ contains a combination of four form factors
that must be calculated nonperturbatively. At zero recoil ${\cal
G}(1)=1$, and ${\cal F}_{B\rightarrow D^*}(1)$ reduces to a single
form factor, $h_{A_1}(1)$.  Given $h_{A_1}(1)$, the measured decay rate determines
$|V_{cb}|$.

The quantity $h_{A_1}$ is a form factor of the axial vector
current,
\bea \langle D^*(v,\epsilon')|{\cal A}^\mu|\overline{B}(v)\rangle &=& i
\sqrt{2m_B 2m_{D^*}} \ \overline{\epsilon'}^\mu h_{A_1}(1), \eea
\noindent where ${\cal A}^\mu$ is the continuum axial-vector current and $\epsilon'$ is the polarization vector of the $D^*$.  Heavy-quark symmetry plays a useful role in constraining
$h_{A_1}(1)$, leading to the heavy-quark expansion
\cite{Falk:1992wt,Mannel:1994kv}
\bea 
h_{A_1}(1) &=&
\eta_A\left[1-\frac{\ell_V}{(2m_c)^2}+\frac{2\ell_A}{2m_c
2m_b}-\frac{\ell_P}{(2m_b)^2}\right], \label{eq:hA1}
\eea
up to order $1/m_Q^2$, and where $\eta_A$ is a factor that matches
heavy-quark effective theory (HQET) to QCD
\cite{Czarnecki:1996gu,Czarnecki:1997cf}. The $\ell$'s are long
distance matrix elements of the HQET. Heavy-quark symmetry forbids
terms of order $1/m_Q$ at zero recoil \cite{Luke:1990eg}, and various methods have
been used to compute the size of the $1/m_Q^2$ coefficients,
including quenched lattice QCD \cite{Hashimoto:2001nb}.  

The earlier work by Hashimoto \etal\  \cite{Hashimoto:2001nb} used 
three double ratios in
order to obtain separately each of the three $1/m_Q^2$
coefficients in Eq.~(\ref{eq:hA1}). These three double ratios also
determine three out of the four coefficients appearing at
$1/m_Q^3$ in the heavy-quark expansion.  It was shown in
Ref.~\cite{Kronfeld:2000ck} that, for the Fermilab method matched
to tree level in $\alpha_s$ and to next-to-leading order in HQET, the
leading discretization errors for the double ratios for this
quantity are of order 
$\alpha_s(\overline{\Lambda}/2m_Q)^2 f_{\mathcal{B}}(am_Q)$ and
$(\overline{\Lambda}/2m_Q)^3 f_i(am_Q)$, where $\overline{\Lambda}$ is a
QCD scale stemming from the light degrees of freedom, such as that 
appearing in the HQET expansion for the heavy-light meson mass,
$m_M=m_Q+\overline{\Lambda} +\cdots$.  
The functions $f_i(am_Q)$ are coefficients depending on $am_Q$ and $\alpha_s$, 
but not on $\overline{\Lambda}$.  
When $am_Q\sim1$, the $f_i(am_Q)$ are of order one; when $am_Q\ll 1$,
they go like a power of $am_Q$, such that the continuum limit is obtained.  The powers of 2 are combinatoric factors.

As discussed in
Ref.~\cite{Hashimoto:2001nb}, all uncertainties in the double ratios
${\cal R}$ used in that work scale as ${\cal R}-1$ rather than as
${\cal R}$.  Statistical errors in the numerator and denominator
are highly correlated and largely cancel in these double ratios.  Also, most
of the normalization uncertainty in the lattice currents cancels,
leaving a normalization factor close to one which can be computed
reliably in perturbation theory.  Finally, the quenching error,
relevant to Ref.~\cite{Hashimoto:2001nb} but not to the present
unquenched calculation, scales as ${\cal R}-1$ rather than as
${\cal R}$. This scaling of the error occurs because the
double ratios constructed in Ref.~\cite{Hashimoto:2001nb} become
the identity in the limit of equal bottom and charm quark masses.

In the calculation reported here, the form factor $h_{A_1}(1)$ is
computed more directly using only one double ratio,
\bea
\label{eq:doubleR} {\cal R}_{A_1}=\frac{\langle
D^*|\overline{c}\gamma_j \gamma_5 b|\overline{B}\rangle \langle
\overline{B}|\overline{b}\gamma_j \gamma_5 c|D^*\rangle}{\langle
D^*|\overline{c}\gamma_4 c|D^*\rangle \langle
\overline{B}|\overline{b}\gamma_4 b|\overline{B}\rangle} =
\left|h_{A_1}(1)\right|^2, 
\eea
which is exact to all orders in the heavy-quark
expansion in the continuum.\footnote{Note that 
the notation ${\cal R}_{A_1}$ stands for
a different double ratio in Ref.~\cite{Hashimoto:2001nb}.}  The lattice
approximation to this ratio still has discretization errors that are
suppressed by inverse powers of heavy-quark masses
[$\alpha_s(\overline{\Lambda}/2m_Q)^2$ and $(\overline{\Lambda}/2m_Q)^3$],
but which again vanish in the continuum limit.  
The errors in the ratio introduced in Eq.~(\ref{eq:doubleR}) do not
scale rigorously
as ${\cal R}_{A_1}-1$ because ${\cal R}_{A_1}$ is not one in the limit
of equal bottom and charm quark masses.
Nevertheless, this double ratio still
retains the desirable features of the previous double ratios, i.e.,
large statistical error cancellations and the cancellation of most
of the lattice current renormalization.  Because the quenching
error has been eliminated, 
the rigorous
scaling of all the errors as ${\cal R}-1$, including the quenching
error, is no longer crucial. The more direct method
introduced here has the significant advantage that extracting
coefficients from fits to HQET expressions as a function of
heavy-quark masses is not necessary, and no error is introduced
from truncating the heavy-quark expansion to a fixed order in
$1/m_Q^n$.
In short, for an unquenched QCD calculation, the method using
Eq.~(\ref{eq:doubleR}) gives a smaller total error than the method used in
Ref.~\cite{Hashimoto:2001nb} for a fixed amount of computer time .

The currents of lattice gauge theory must be matched to the 
normalization of the continuum to obtain~$\mathcal{R}_{A_1}$.
The matching factors mostly cancel in the double ratio~\cite{Harada:2001fi, Harada:2001fj}, 
leaving $h_{A_1}(1) = \sqrt{\mathcal{R}_{A_1}}=\rho\sqrt{R_{A_1}}$, 
where $R_{A_1}$ is the lattice double ratio and $\rho$, the ratio of 
matching factors, is very close to~1.
(For the remainder of this paper we shall use
the convention that a script
letter corresponds to a continuum quantity, while a non-script letter
corresponds to a lattice quantity.)  This $\rho$ factor has
been calculated to one-loop order in perturbative QCD, and 
is found to contribute less than a $0.5\%$ correction.  
We have exploited the $\rho$ factors to implement a blind analysis.
Two of us involved in the perturbative calculation applied a common
multiplicative offset to the $\rho$ factors needed to obtain
$h_{A_1}(1)$ at different lattice spacings.
This offset was not disclosed to the rest of us until the procedure for determining the systematic
error budget for the rest of the analysis had been finalized. 

The unquenched MILC configurations generated with 2+1 flavors of
improved staggered fermions make use of the fourth-root procedure for
eliminating the unwanted four-fold degeneracy of 
staggered quarks.  
At non-zero lattice spacing, this procedure has small violations of 
unitarity \cite{Prelovsek:2005rf, Bernard:2006gj, Bernard:2006zw, Bernard:2007qf, Aubin:2008wk} and 
locality \cite{Bernard:2006ee}.
Nevertheless, a careful treatment of the continuum limit, in
which all assumptions are made explicit, argues that lattice QCD with
rooted staggered quarks reproduces the desired local theory of QCD as
$a\to0$ \cite{Shamir:2004zc, Shamir:2006nj}.
When coupled with other analytical and numerical
evidence (see Refs.~\cite{Durr:2005ax,Sharpe:2006re,Kronfeld:2007ek} for reviews),
this gives us confidence that the rooting procedure is indeed correct
in the continuum limit.

The outline of the rest of this paper is as follows: Section II
describes the details of the lattice simulation.  Section III discusses
the fits to the double ratios accounting for oscillating opposite-parity
states.  Section IV summarizes the lattice perturbation theory
calculation of the $\rho$ factor.  Section V introduces the rooted
staggered chiral perturbation theory formalism and expressions used in
the chiral extrapolations.  Section VI then discusses our treatment of
the chiral extrapolation and introduces our approach for disentangling
heavy and light-quark discretization effects.  Section VII provides a
detailed discussion of our systematic errors, and we conclude in Section
VIII.

\section{Lattice calculation}

The lattice calculation was done on the MILC ensembles at three lattice
spacings with $a \approx 0.15$, $0.125$, and $0.09$~fm; these ensembles have
an $O(a^2)$ Symanzik improved gauge action and 2+1 flavors of ``AsqTad''
improved staggered sea quarks \cite{Blum:1996uf, Orginos:1998ue, Lagae:1998pe, Lepage:1998vj, Orginos:1999cr, Bernard:1999xx}.  The parameters for the MILC lattices used
in this calculation are shown in Table~\ref{tab:params}.  We have
several light masses at both full QCD and partially-quenched points
($m_{\rm val}\neq m_{\rm sea}$), and our light quark masses range between
$m_{s}/10$ and $m_{s}/2$.  Table~\ref{tab:params2} shows the valence
masses computed on each ensemble.  In this work we follow the
notation \cite{Aubin:2004fs} where $m_{s}$ is the physical strange quark mass, $\hat{m}$ is
the average $u$-$d$ quark mass, and 
$\hat{m}'$, $m'_s$ indicate the nominal values used in simulations.
In practice, the MILC ensembles choose $m'_s$ within 10--30\%
of $m_s$ and a range of $\hat{m}'$ to enable a chiral 
extrapolation.

The heavy quarks are computed using the Sheikholeslami-Wohlert (SW)
``clover'' action \cite{Sheikholeslami:1985ij} with the Fermilab
interpretation via HQET \cite{ElKhadra:1996mp}. 
The SW action includes a dimension-five interaction with a coupling
$c_{\textrm{SW}}$ that has been adjusted to the value $u_0^{-3}$
suggested by tadpole-improved, tree-level perturbation theory
\cite{Lepage:1992xa}.  The value of $u_0$ is calculated either
from the plaquette ($a\approx 0.15$ fm and $a\approx 0.09$ fm), or
from the Landau link ($a \approx 0.12$ fm). The adjustment of
$c_{\textrm{SW}}$ 
is needed to normalize the heavy quark's chromomagnetic moment correctly 
\cite{ElKhadra:1996mp}.

\begin{table}
\begin{center}
\caption{Parameters of the simulations.  The columns from left to right are the approximate lattice spacing in fm, the sea quark masses $a\hat{m}'/am'_s$, the linear spatial dimension of the lattice ensemble in fm, the dimensionless factor $m_\pi L$ ($m_\pi$ corresponds to the taste-pseudoscalar pion composed of light sea quarks), the gauge coupling, the dimensions of the lattice in lattice units, the number of configurations used for this analysis, the bare hopping parameter used for the bottom quark, the bare hopping parameter used for the charm quark, and the clover term $c_{SW}$ used for both bottom and charm quarks.}
 \label{tab:params}
\begin{tabular}{cccccccccc}
  \hline \hline
  $a$(fm) & $a\hat{m}' / am'_s$ & $L$(fm) & \ $m_\pi L$ \ & $10/g^2$ & Volume & $\#$ Configs & $\kappa_b$ & $\kappa_c$ & $c_{SW}$\\
  \hline
  $0.15$  & $0.0194/0.0484$ & 2.4 & 5.5 & \ 6.586 \ & $16^3\times 48$ & 628 & 0.076 \ & \ 0.122 & \ 1.5673\\
  $0.15$  & $0.0097/0.0484$ & 2.4 & 3.9 & 6.572 & $16^3 \times 48$ & 628 & 0.076 \ & \ 0.122 & \ 1.5673\\
  \hline
  $0.12$ & $0.02/0.05$ & 2.4 & 6.2 & 6.79 & $20^3\times 64$ & 460 & 0.086 & \ 0.122 & \ 1.72\\
  $0.12$ & $0.01/0.05$ & 2.4 & 4.5 & 6.76 & $20^3\times 64$ & 592 & 0.086 & \ 0.122 & \ 1.72\\
  $0.12$ & $0.007/0.05$ & 2.4 & 3.8 & 6.76 & $20^3\times 64$ & 836 & 0.086 & \ 0.122 & \ 1.72 \\
  $0.12$ & $0.005/0.05$ & 2.9 & 3.8 & 6.76 & $24^3\times 64$ & 528 & 0.086 & \ 0.122 & \ 1.72\\
  \hline
  $0.09$ & $0.0124/0.031$ & 2.4 & 5.8 & 7.11 & $28^3\times 96$ & 516 & 0.0923 & \ 0.127 & \ 1.476\\
  $0.09$ & $0.0062/0.031$ & 2.4 & 4.1 & 7.09 & $28^3\times 96$ & 556 & 0.0923 & \ 0.127 & \ 1.476\\
  $0.09$ & $0.0031/0.031$ & 3.4 & 4.2 & 7.08 & $40^3\times 96$ & 504 & 0.0923 & \ 0.127 & \ 1.476 \\
  \hline \hline
\end{tabular}
\end{center}
\end{table}

\begin{table}
\begin{center}
\caption{Valence masses used in the simulations.  The columns from left to right are the approximate lattice spacing in fm, the sea quark masses $a\hat{m}'/am'_s$ identifying the gauge ensemble, and the valence masses computed on that ensemble.}
 \label{tab:params2}
\begin{tabular}{ccc}
  \hline \hline
  $a$(fm) & $a\hat{m}' / am'_s$ & $am_x$\\
  \hline
  $ \ \ \approx 0.15 \ \ $ & \ \ $0.0194/0.0484 \ \ $ & 0.0194 \\
  $\approx 0.15$ & \ \ $0.0097/0.0484 \  \ $ & \ \ 0.0097, 0.0194 \ \ \\
  \hline
  $\approx 0.12$ & $0.02/0.05$ & 0.02 \\
  $\approx 0.12$ & $0.01/0.05$ & 0.01, 0.02\\
  $\approx 0.12$ & $0.007/0.05$ &  0.007, 0.02\\
  $\approx 0.12$ & $0.005/0.05$ & 0.005, 0.02\\
  \hline
  $\approx 0.09$ & $0.0124/0.031$ & 0.0124\\
  $\approx 0.09$ & $0.0062/0.031$ & 0.0062, 0.0124\\
  $\approx 0.09$ & $0.0031/0.031$ & 0.0031, 0.0124\\
  \hline \hline
\end{tabular}
\end{center}
\end{table}

The tadpole-improved bare quark mass for SW quarks is given by
\bea 
am_0 =
\frac{1}{u_0}\left(\frac{1}{2\kappa}-\frac{1}{2\kappa_{\rm crit}}\right),
\eea
where tuning the parameter $\kappa$ to the critical
quark hopping parameter $\kappa_{\rm crit}$ would lead to a massless
pion.  The spin averaged $B_s$ and $D_s$ kinetic masses 
are computed on
a subset of the ensembles in order to tune the bare $\kappa$ values for
bottom and charm (and hence the corresponding bare quark masses)
to their physical values.  These tuned values were then used in
the $B\to D^* \ell \nu$ form-factor production run.

The relative lattice scale is determined by
calculating $r_1/a$ on each ensemble, where $r_1$ is related to the
force between static quarks by $r_1^2F(r_1)=1.0$
\cite{Sommer:1993ce,Bernard:2000gd}.
To avoid introducing implicit dependence on $\hat{m}'$, $m'_s$ via
$r_1(\hat{m}',m_s',g^2)$ (where, as above, primes denote simulation
masses), we interpolate in $m_s'$ and extrapolate in $\hat{m}'$ to obtain
$r_1(\hat{m},m_s,g^2)/a$ at the physical masses.
We then convert from lattice units to $r_1$ units with
$r_1(\hat{m},m_s,g^2)/a$.
Below we shall call this procedure the mass-independent determination 
of~$r_1$.

In order to fix the absolute lattice scale, one must compute a physical
quantity that can be compared directly to experiment; we use the
$\Upsilon$ 2S--1S splitting \cite{Gray:2005ad} and the most recent MILC
determination of $f_\pi$ \cite{Bernard:2007ps}.  The difference between
these determinations results in a systematic error that turns out to be
much smaller than our other systematics.  When the $\Upsilon$ scale
determination is combined with the continuum extrapolated $r_1$ value
at physical quark masses, a value $r_1^{\rm phys}=0.318(7)$ fm
\cite{Bernard:2005ei} is obtained.  The $f_\pi$ determination is
$r_1^{\rm phys}=0.3108(15)(^{+26}_{-79})$ fm \cite{Bernard:2007ps}.  Given
$r_1^{\rm phys}$, it is then straightforward to convert quantities measured
in $r_1$ units to physical units.

The dependence on the lattice spacing $a$ is mild in this
analysis. Since $a$ only enters the calculation through the
adjustment of the heavy and light quark masses, the dependence of $h_{A_1}(1)$ on
$a$ is small.  Staggered chiral perturbation theory indicates that the
$a$ dependence coming from staggered quark discretization effects is
small \cite{Laiho:2005ue}, and this is consistent with the
simulation data. 

In this work, we construct lattice currents as in
Ref.~\cite{ElKhadra:1996mp},
\bea
J_\mu^{hh'}=\sqrt{Z^{hh}_{V_4}Z^{h'h'}_{V_4}}\bar{\Psi}_h
\Gamma_\mu \Psi_{h'}, 
\label{eq:current}
\eea
where $\Gamma_\mu$ is either the vector ($i\gamma^\mu$)
or axial-vector ($i\gamma^\mu\gamma_5$) current.  The rotated
field $\Psi_h$ is defined by
\bea
\Psi_h = (1+a d_1 \mbox{\boldmath$\gamma$} \cdot {\bf
D}_{\textrm{lat}})\psi_h, 
\eea
where $\psi_h$ is the (heavy) lattice quark field in the
SW action. $\textbf{D}_{\textrm{lat}}$ is the symmetric,
nearest-neighbor, covariant difference operator; the tree-level
improvement coefficient is
\bea 
d_1=\frac{1}{u_0}\left(\frac{1}{2+m_0 a}-\frac{1}{2(1+m_0 a)}\right).
\eea
In Eq.~(\ref{eq:current}) we choose to normalize the current
by the factors of $Z^{hh}_{V_4}$ ($h=c,b$) since
even for massive quarks they are easy to compute non-perturbatively.
The continuum current is related to the lattice current by
\bea {\cal J}_\mu^{hh'} = \rho_{J_\Gamma} J^{hh'}_\mu
\eea
\noindent up to discretization effects, where 
\bea\label{eq:rho_ratio} \rho^2_{J_\Gamma} =
\frac{Z^{bc}_{J_\Gamma}Z^{cb}_{J_\Gamma}}{Z^{cc}_{V_4}Z^{bb}_{V_4}},
\eea
and the
matching factors $Z^{hh'}_{J_\Gamma}$'s are defined
in Ref.~\cite{Harada:2001fj}.
%
%
Note that the factor $\sqrt{Z^{bb}_{V_4}Z^{cc}_{V_4}}$ multiplying the
lattice current in Eq.~(\ref{eq:current}) cancels in the double ratio by
design, leaving only the $\rho$ factor, which is close to one and can be
computed reliably using perturbation theory.  The perturbative calculation of $\rho_{J_\Gamma}$ is described in more detail in Section~IV.

Interpolating operators are constructed from four-component heavy
quarks and staggered quarks as follows.
Let
\bea
{\cal O}_{D^*_j}(x) & = & \bar{\chi}(x)\Omega^\dagger(x)i\gamma_j\psi_c(x), \\
{\cal O}^\dagger_B(x) & = & \bar{\psi}_b(x)\gamma_5\Omega(x)\chi(x),
\eea
where $\chi$ is the one-component field in the staggered-quark action,
and
\begin{equation}
\Omega(x)=\gamma_1^{x_1/a}\gamma_2^{x_2/a}\gamma_3^{x_3/a}\gamma_4^{x_4/a}.
\end{equation}
The left (right) index of $\Omega^\dagger$ ($\Omega$) can be left as a
free taste index~\cite{Kronfeld:2007ek} or $\chi$ can be promoted to a
four-component naive-quark field to contract all
indices~\cite{Wingate:2003nn}.
The resulting correlation functions are the same if the initial and
final taste indices are set equal and then summed.
The same kinds of operators have been used in previous calculations~\cite{Okamoto:2004xg, Aubin:2004ej, Aubin:2005ar}.

Lattice matrix elements are obtained from three-point correlation
functions.  The three-point correlation functions needed for the
$B\to D^*$ transition at zero-recoil are
\bea\label{eq:corr} C^{B\to
D^*}(t_i,t_s,t_f)=\sum_{\textbf{x},\textbf{y}}\langle 0|{\cal
O}_{D^*}(\textbf{x},t_f)\overline{\Psi}_c\gamma_j\gamma_5
\Psi_b(\textbf{y},t_s){\cal O}^\dag_B(\textbf{0},t_i)| 0\rangle,
\\ C^{B\to B}(t_i,t_s,t_f)=\sum_{\textbf{x},\textbf{y}}\langle
0|{\cal O}_{B}(\textbf{x},t_f)\overline{\Psi}_b\gamma_4
\Psi_b(\textbf{y},t_s){\cal O}^\dag_B(\textbf{0},t_i)| 0\rangle,
\\ \label{eq:corr3} C^{D^*\to D^*}(t_i,t_s,t_f)=\sum_{\textbf{x},\textbf{y}}\langle
0|{\cal O}_{D^*}(\textbf{x},t_f)\overline{\Psi}_c\gamma_4
\Psi_c(\textbf{y},t_s){\cal O}^\dag_{D^*}(\textbf{0},t_i)|
0\rangle. \eea
\noindent  In $C^{B\to D^*}$ the
polarization of the $D^*$ lies along spatial direction~$j$.
If the source-sink separation is large enough then we can arrange for both
$t_s-t_i$ and $t_f-t_s$ to be large so that the lowest-lying state dominates.  Then
\bea
\label{eq:grndstate} C^{B\to D^*}(t_i,t_s,t_f) &=&
{\cal Z}^{\frac12}_{D^*}{\cal Z}^{\frac12}_B \frac{\langle D^*|\overline{\Psi}_c
\gamma_j \gamma_5 \Psi_b|B \rangle}{\sqrt{2m_{D^*}}\sqrt{2m_B}} \
e^{-m_B (t_s-t_i)}e^{-m_{D^*}(t_f-t_s)} + ..., 
\eea
where $m_B$ and $m_{D^*}$ are the masses of the $B$ and
$D^*$ mesons and 
${\cal Z}_H=|\langle 0|\mathcal{O}_H|H\rangle|^2$.

In practice, the meson source and sink are held at fixed $t_i=0$ and
$t_f=T$, while the operator time $t_s=t$ is varied over all times in
between.  Using the correlators defined in
Eqs.~(\ref{eq:corr}-\ref{eq:corr3}) we form the double ratio
\bea
\label{eq:DR} R_{A_1}(t) = \frac{C^{B\to D^*}(0,t,T)C^{D^*\to
B}(0,t,T)}{C^{D^*\to D^*}(0,t,T)C^{B\to B}(0,t,T)}. 
\eea
All convention-dependent normalization factors, including the factors of
$\sqrt{{\cal Z}_H/2m_H},$ cancel in the double ratio.  
In the window of time separations where the ground state dominates, a plateau should be visible, and
the lattice ratio is simply related to the
continuum ratio ${\cal R}_{A_1}$ by a renormalization factor
\bea
\label{eq:rhodef} \rho_{A_1} \sqrt{R_{A_1}}=\sqrt{{\cal
R}_{A_1}}=h_{A_1}(1), 
\eea
with $\rho_{A_1}$ as in Eq.~(\ref{eq:rho_ratio}).
The right-hand side of Eq.~(\ref{eq:grndstate}) is the first term
in a series, with additional terms for each radial excitation,
including opposite-parity states 
that arise with staggered quarks.
Eliminating the opposite-parity states requires some care, and this is
discussed in detail in the next section.  In order to isolate the
lowest-lying states we have chosen creation and annihilation
operators, ${\cal O}^\dag_B$ and ${\cal O}_{D^*}$, that have a
large overlap with the desired state. This was done by smearing
the heavy quark and anti-quark propagator sources with 1S
Coulomb-gauge wave-functions.

\section{Fitting and opposite-parity states}

Extracting correlation functions of operators with staggered quarks
presents an extra complication because the
contributions of opposite-parity states introduce oscillations in time
into the correlator fits \cite{Wingate:2003nn}.  
Three-point functions obey the functional form
\bea\label{eq:oscil} C^{X\to Y}(0,t,T) =
\sum_{k=0} \sum_{\ell=0} (-1)^{kt}(-1)^{\ell(T-t)}A_{\ell
k}e^{-m^{(k)}_{X}t}e^{-m^{(\ell)}_{Y}(T-t)}. \eea
\noindent For odd $k$ and $\ell$ the excited state contributions
change sign as the position of the operator varies by one time
slice. Although they are exponentially suppressed, the parity
partners of the heavy-light mesons are not that much heavier than
the ground states in which we are interested, so the oscillations
can be significant at the source-sink separations typical of our
calculations. These separations cannot be too large because
of the rapid decrease of the signal due to the presence of the heavy quark.

Although one can fit a given three-point correlator to
Eq.~(\ref{eq:oscil}), in the calculation of $h_{A_1}(1)$ we use double ratios in which numerator and denominator are so similar
that most of the fitting systematics cancel, and it is convenient to
preserve this simplifying feature.  We do this by forming a suitable
average over correlator ratios with different (even and odd) source-sink
separations.  It turns out that the amplitudes of the oscillating states
in $B \to D^*$ correlation functions are much smaller than they are in
many other heavy-light transitions \cite{Dalgic:2006dt, Evans:2006zz}, and that the oscillating states in
$B\to D^*$ are barely visible at the present level of statistics.  Even
so, we introduce an average that reduces them still further, to the
point where they are negligible.

Although we shall take the average of the
double ratio, let us first examine the average of an individual
three-point function.  Expanding Eq.~(\ref{eq:oscil}) so that it
includes the ground state and the first oscillating state, we have
\bea
\label{eq:Cexpand} C^{X\to Y}(0,t,T)&=& A^{X\to Y}_{00}e^{-m_X t - m_Y(T-t)} +
(-1)^{T-t} A^{X\to Y}_{01}e^{-m_X t-m_Y'(T-t)} \nonumber \\ &&
+(-1)^t A^{X\to Y}_{10} e^{-m_X' t -m_Y(T-t)}+(-1)^T A^{X\to
Y}_{11} e^{-m_X't-m_Y'(T-t)}+... \nonumber \\ && =A^{X\to
Y}_{00}e^{-m_X t-m_Y(T-t)}\left[1+c^{X\to
Y}(0,t,T)+...\right],
\eea
where in the last line we have pulled out the ground
state amplitude and exponential dependence.  The function $c^{X\to Y}(0,t,T)$ is defined
\bea
\label{eq:c} c^{X\to Y}(0,t,T) &\equiv & \frac{A^{X\to
Y}_{01}}{A^{X\to Y}_{00}}(-1)^{T-t}e^{-\Delta m_Y(T-t)} +
\frac{A^{X\to Y}_{10}}{A^{X\to Y}_{00}}(-1)^t e^{-\Delta m_X t}
\nonumber \\ && + \frac{A^{X\to Y}_{11}}{A^{X\to Y}_{00}}(-1)^T
e^{-\Delta m_X t -\Delta m_Y(T-t)},
\eea
where $\Delta m_{X,Y}=m'_{X,Y}-m_{X,Y}$ 
is the splitting between the 
lowest-lying desired-parity state and the lowest-lying wrong-parity 
state.
Note that the first two terms produce oscillations as
the position of the operator is varied over the time extent of the
lattice.  The third term, however, changes sign only when the
total source-sink separation is varied.  It is this term that our
average is designed to suppress, since it will not be as clearly
visible in the $t$ dependence of the lattice data as 
those that oscillate in~$t$.

We define the average to be
\bea\label{eq:avgdef} \overline{C}^{X\to Y}(0,t,T) &\equiv &
\frac{1}{2}C^{X\to Y}(0,t,T) + \frac{1}{4}C^{X\to Y}(0,t,T+1)
\nonumber \\ && + \frac{1}{4}C^{X\to Y}(0,t+1,T+1).
\eea
Substituting the expression for $C^{X\to Y}(0,t,T)$ from Eq.~(\ref{eq:Cexpand}) into this definition gives
\bea\label{eq:avgdef2} \overline{C}^{X\to Y}(0,t,T)= A^{X\to Y}_{00}e^{-m_X t-m_Y(T-t)}\left[1+\overline{c}^{X\to
Y}(0,t,T)+...\right], \eea
\noindent where the function $\overline{c}^{X\to Y}$ is
\bea\label{eq:average} \overline{c}^{X\to Y}(0,t,T) &\equiv &
\frac{A^{X\to Y}_{01}}{A^{X\to Y}_{00}}(-1)^{T-t}e^{-\Delta
m_Y(T-t)}\left[\frac{1}{2}+\frac{1}{4}(1-e^{-\Delta m_Y})\right]
\nonumber
\\ && + \frac{A^{X\to Y}_{10}}{A^{X\to Y}_{00}}(-1)^t e^{-\Delta m_X
t}\left[\frac{1}{2}+\frac{1}{4}(1-e^{-\Delta m_X})\right]
\nonumber
\\ && + \frac{A^{X\to Y}_{11}}{A^{X\to Y}_{00}}(-1)^T e^{-\Delta
m_X t -\Delta m_Y(T-t)}\left[\frac{1}{2}-\frac{1}{4}(e^{-\Delta
m_Y}+e^{-\Delta m_X})\right]. \eea
\noindent Note that Eq.~(\ref{eq:average}) has the same
exponential time dependence as Eq.~(\ref{eq:c}), but with the size
of the amplitudes reduced by the factors in square
brackets.  Thus, the average is equivalent to a smearing that reduces
the oscillating state amplitudes. 
It is possible to compute the $\Delta m_X$
precisely from fits to
two-point correlators.  We find values between about $0.2$ and $0.4$ in
lattice units.  Given these values, 
the first two factors in
brackets reduce their respective amplitudes by
approximately a factor of 2, and the 
targeted, non-oscillating term
is reduced by a factor of $\sim 6$--$10$.

Specializing to the $B\to D^*$ case, consider the double ratio
\bea\label{eq:DRexpand} R_{A_1}(0,t,T) &=& \frac{A^{B\to D^*}_{00}A^{D^*\to
B}_{00}}{A^{D^*\to D^*}_{00}A^{B\to B}_{00}}\left[1+c^{B\to
D^*}(0,t,T)+c^{D^*\to B}(0,t,T)\right. \nonumber \\ && \left. -
c^{D^*\to D^*}(0,t,T) - c^{B\to B}(0,t,T)+...\right], \eea
\noindent where we have again factored out the ground state
contribution.  Equation~(\ref{eq:DRexpand}) follows from Eq.~(\ref{eq:DR}) treating the $c$'s as small.  Note that the $c$'s are expected to be similar in
numerator and denominator, and to the extent that they are the
same they will cancel in this expression.  Applying the average in
Eq.~(\ref{eq:avgdef}) directly to the double ratio,
\bea\label{eq:avg} \overline{R}(0,t,T) &\equiv & \frac{1}{2}R(0,t,T) +
\frac{1}{4}R(0,t,T+1) \nonumber \\ && + \frac{1}{4}R(0,t+1,T+1),
\eea
\noindent we get
\bea\label{eq:Ra1avg} \overline{R}_{A_1}(0,t,T) &=& \frac{A^{B\to D^*}_{00}A^{D\to
B^*}_{00}}{A^{D\to D^*}_{00}A^{B\to
B^*}_{00}}\left[1+\overline{c}^{B\to D^*}(0,t,T)+\overline{c}^{D\to
B^*}(0,t,T)\right. \nonumber \\ && \left. - \overline{c}^{D\to
D^*}(0,t,T) - \overline{c}^{B\to B^*}(0,t,T)+...\right], \eea
\noindent where each of the oscillating state terms in the
individual three-point functions is suppressed according to
Eq.~(\ref{eq:average}).

\begin{figure}
\begin{center}
 \includegraphics[scale=.55]{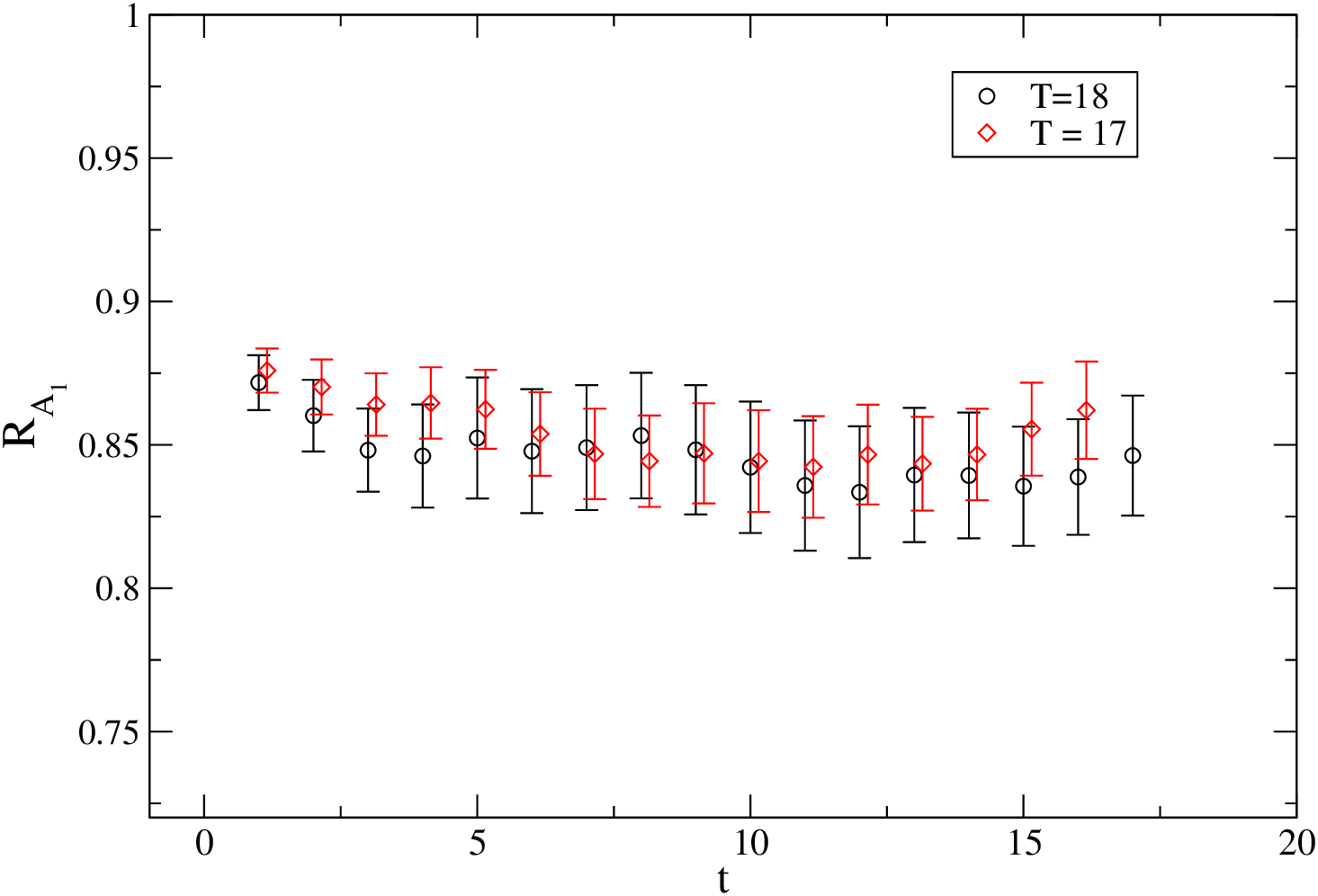}
\end{center}
\caption{Double ratio $R_{A_1}$ on the $a\hat{m}'= 0.0124$ 
fine ($a=0.09$ fm) ensemble. The source was fixed to time slice 0, and the
operator position was varied as a function of time.  Two different
sink points were used with even and odd time separations between
source and sink in order to study the effect of non-oscillating contributions from
wrong parity states.
\label{fig:doubleRevenodd}}
\end{figure}

\begin{figure}
\begin{center}
 \includegraphics[scale=.55]{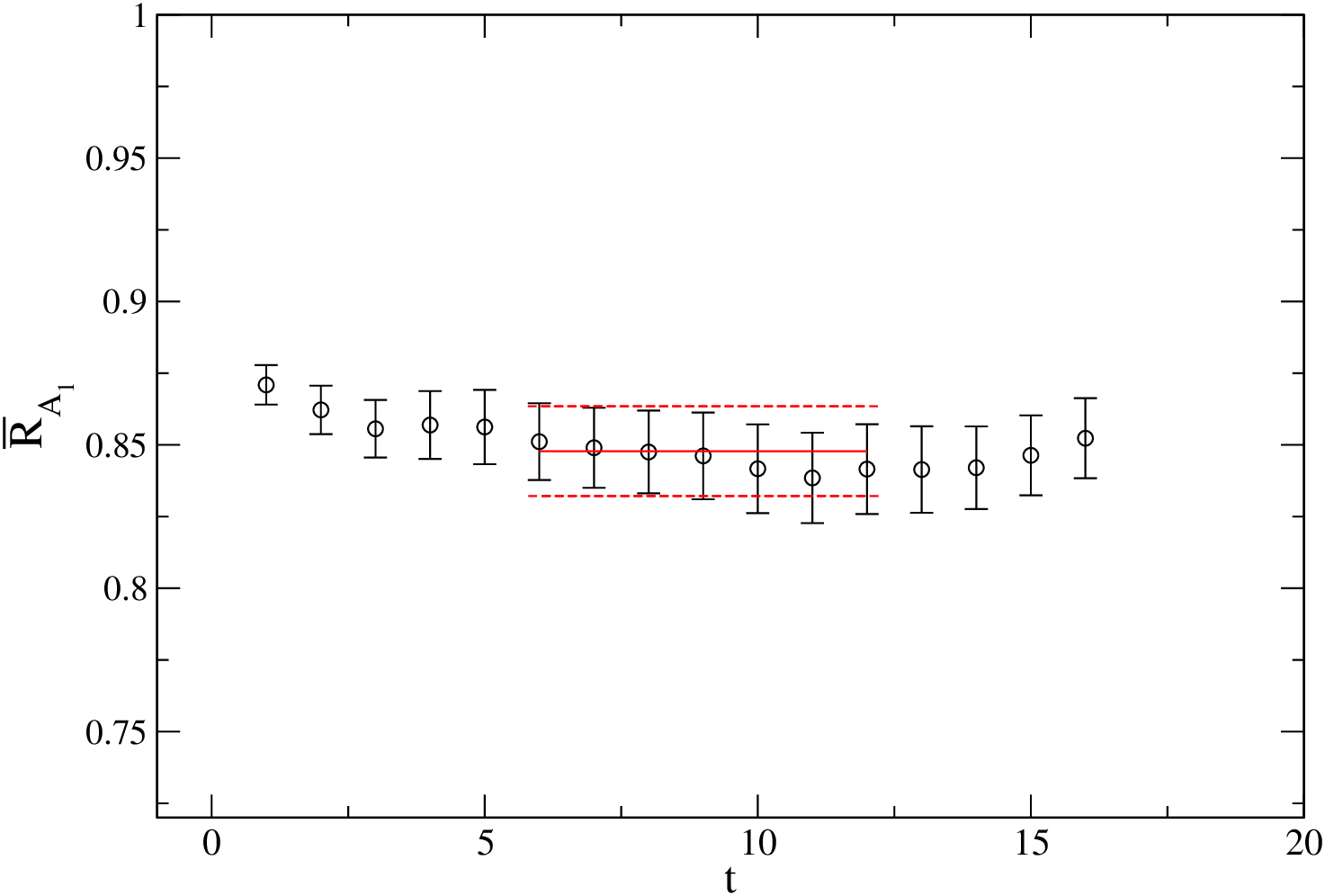}
\end{center}
\caption{Averaged double ratio, $\overline{R}_{A_1}$, of
Eq.~(\ref{eq:avg}) on the $a\hat{m}'= 0.0124$ fine ($a=0.09$ fm) ensemble.  The plateau
fit is shown with 1$\sigma$ error band.
\label{fig:doubleR}}
\end{figure}

Although $\Delta m_B$ and $\Delta m_{D^*}$ can be obtained from fits to the two point correlators, the oscillating state
amplitudes appearing in the three-point correlators must be
determined directly from the three-point correlator data.
Figure~\ref{fig:doubleRevenodd} shows the double ratio
$R_{A_1}$ used to obtain $h_{A_1}(1)$.  
The source is at time slice 0, the
sink is at $T$, and the operator position is varied along $t$. Two
different source-sink separations were generated that differed by
a single time unit at the sink ($T=17,18$). The average of these
two correlators was taken according to Eq.~(\ref{eq:avg}), and
this average was fit (including the full covariance
matrix) to a constant, as shown in Fig.~\ref{fig:doubleR}.
There is no
detectable oscillation even before the average is taken, as can be seen in 
Fig.~\ref{fig:doubleRevenodd}; according
to Eq.~(\ref{eq:average}) the oscillating contributions are
reduced even further in the average so that their systematic
errors can be safely neglected.

\section{Perturbation theory}

Lattice perturbation theory is needed in order to calculate the
short-distance coefficient $\rho_{A_1}$ defined in
Eq.~(\ref{eq:rho_ratio}).  Although naive lattice perturbation theory
appears to converge slowly, the two main causes have been
identified \cite{Lepage:1992xa}:  the bare gauge coupling is a
poor expansion parameter, and coefficients are large when tadpole
diagrams occur.  If a renormalized coupling is used as an
expansion parameter, and one computes only those quantities for
which the tadpole diagrams largely cancel, then lattice
perturbation theory seems to converge as well as perturbation
theory in continuum~QCD.

Only the vertex correction contributes to the $\rho$ factor, as
the wave-function renormalization (including all tadpoles) cancels
by construction.  Even the vertex correction partially cancels, and the
one-loop coefficient is found to be small.  The perturbative corrections
to the $\rho$ factor can be written as
\bea\label{eq:rho} \rho^{hh'}_{J_\Gamma}\equiv
\frac{Z^{hh'}_{J_\Gamma}}{\sqrt{Z^{hh}_{V_4}Z^{h'h'}_{V_4}}} = 1 +
\alpha_V(q^*)4\pi\rho^{hh'[1]}_{J_\Gamma}+...\eea
\noindent where $\rho^{hh'[1]}_{J_\Gamma}$ is the coefficient of
the one-loop correction, and the coupling $\alpha_V$ is the
renormalized strong coupling constant in the V-scheme
\cite{Lepage:1992xa, Brodsky:1982gc}, which is based on the static-quark potential. The coupling is determined
following the procedure of Ref.~\cite{Mason:2005zx}. The scale $q^*$ of
the running coupling $\alpha_V(q^*)$ should be chosen to be the
typical momentum of a gluon in the loop.  
%
%
%
%
A prescription for calculating this scale was introduced by Brodsky, Lepage, and Mackenzie (BLM)~\cite{Lepage:1992xa, Brodsky:1982gc}.  They define $q^*$ by
\begin{equation}\label{eq:q*}
\ln(q^{*2}) = \frac{\int  d^4q \,  f(q) \ln(q^2)} { \int  d^4q  \, f(q)},
\end{equation}
where $f(q)$ is the one-loop integrand and the numerator is the first
log moment.  This prescription was extended by Hornbostel, Lepage, and
Morningstar (HLM)~\cite{Hornbostel:2002af} to cases where the one-loop
contribution is anomalously small leading to a break down of
Eq.~(\ref{eq:q*}).  The HLM prescription for $q^*$ takes into account
two-loop contributions to the gluon propagator via the inclusion of
second log moments.  Since we do encounter anomalously small one-loop
corrections in $\rho_{A_1}$, the HLM prescription was used to determine
$q^*$.  
Results for $q^*_{HLM}$ and $\rho_{A_1}$ needed 
for this calculation are given in
Table~\ref{tab:rho}.
\begin{table}
\begin{center}
\caption{Computed values of $\rho_{A_1}$ in the HLM
prescription~\cite{Hornbostel:2002af} .  The first three columns label
each ensemble with
the approximate lattice
spacing in fm, the light sea quark mass $a\hat{m}'$, and the strange
quark mass $am_s'$.  The fourth column is $aq^*_{HLM}$, where the
error is calculated using the statistical error from VEGAS for the 0th,
1st, and 2nd moments of the one-loop integrals.  The fifth column is
$\rho_{A_1}$ on that ensemble, and the errors are the statistical errors
from the VEGAS evaluation, including the one-loop coefficients and
$q_{HLM}^*$.}
 \label{tab:rho}
\begin{tabular}{ccccc}
\hline \hline
 $a$ (fm) & $a\hat{m}'$ & $am_s'$ & $aq^*_{HLM}$ & $\rho_{A_1}$ \\
  \hline
  $0.15$  & 0.0194 \ \ & 0.0484 & \ \ 2.03(10) \ \ & 0.9966(2)\\
  $0.15$  & 0.0097 \ \ & 0.0484 & 2.03(10) & 0.9966(2)\\
  $0.12$ & 0.02 & 0.05  & 1.96(10) & 0.9964(2)\\
   $0.12$ & 0.01 & 0.05  & 1.96(10) & 0.9964(2)\\
    $0.12$ & 0.007 & 0.05  & 1.96(10) & 0.9964(2)\\
 $0.12$ & 0.005 & 0.05  & 1.96(10) & 0.9964(2)\\
  $0.09$ & 0.0124 & 0.031  & 2.98(14) & 1.00298(9) \\
  $0.09$ & 0.0062 & 0.031  & 2.98(14) & 1.00300(9) \\
   $0.09$ & 0.0031 & 0.031  & 2.98(14) & 1.00301(9) \\
  \hline
\end{tabular}
\end{center}
\end{table}
The $\rho$ factor varies somewhat as a function of lattice
spacing, and is even slightly different from ensemble to ensemble at the
same nominal lattice spacing, due to the slightly different $\beta$
values used to generate the gauge fields.


The calculation of $\rho_{A_1}$ is described in
Refs.~\cite{ElKhadra:2007qe, ElKhadra:2008}.  It uses automated
perturbation theory techniques to generate the Feynman rules and VEGAS
\cite{Lepage:1977sw} for the numerical integration of the loop
integrals.  As a check, it was verified that this calculation reproduces
known results for the heavy-heavy currents with the Wilson plaquette
action \cite{Harada:2001fi} and for the $V_4$ current in the massless
limit with the Symanzik improved gauge action.

As mentioned in the introduction, we have exploited the $\rho$ factor to implement a blind 
analysis.
Two of us applied a multiplicative offset close to~1 to the $\rho$ factor, 
generated with a random key.
The offset was not unlocked until the procedure for determining the systematic errors in the rest of
the analysis had been finalized.

\section{Staggered Chiral Perturbation Theory}

The simulation masses $\hat{m}'_{\rm val}$ and $\hat{m}'_{\rm sea}$ 
(for valence and sea) are all larger than the physical $\hat{m}$.
A controlled chiral extrapolation can be guided by an appropriate chiral
effective theory that includes the effect of staggered-quark
discretization errors.  Rooted staggered chiral perturbation theory  (rS$\chi$PT),
which has been formulated for heavy-light
quantities in Ref.~\cite{Aubin:2005aq}, is such a theory.
In rS$\chi$PT, a replica method is used to take into account the effect
of rooting; this procedure has been justified in
Refs.~\cite{Bernard:2006zw, Bernard:2007ma}.  

Because of taste-symmetry
breaking, the staggered theory has 16 light pseudoscalar mesons instead of 1.  The
tree-level relation for the masses of light staggered mesons in the chiral theory is \cite{Lee:1999zx,Aubin:2003mg}
\bea  
	m_{xy,\Xi}^2 & = & \mu_{0} (m_x + m_y) + a^2 \Delta_\Xi\ ,
\eea
where $m_x$ and $m_y$ are staggered quark masses, $\mu_0$ is the
continuum low-energy constant, and $a^2 \Delta_\Xi$ are the
splittings of the 
16 pions of taste~$\Xi$.
For staggered quarks there
exists a residual SO(4) taste symmetry broken at ${\cal O}(a^2)$, such that
there is
some degeneracy among
the 16 pions \cite{Lee:1999zx}, and the
taste index $\Xi$ runs over the multiplets $P, A, T, V, I$ with
degeneracies 1, 4, 6, 4, 1.  The splitting $a^2 \Delta_P$ vanishes
because there is an exact (non-singlet) lattice axial symmetry.

Schematically, the next-to-leading order (NLO) result for the relevant
form factor is
\bea
\label{eq:NLOsChPT} 
h_{A_1}^{\textrm{NLO}}(1)/\eta_A = 1 + X_A(\Lambda_\chi) + \frac{g^2_{DD^*\pi}}{48\pi^2 f^2}\times\textrm{logs}_{\rm 1-loop}(\Lambda_\chi)
\eea
where $X_A(\Lambda_\chi)$ is a low energy constant of the
chiral effective theory, and is therefore independent of light quark
mass and cancels the chiral scale dependence $\Lambda_\chi$ of the chiral logarithms.
By heavy-quark symmetry, $X_A(\Lambda_\chi)$ is proportional to
$1/m^2_c$ in the heavy-quark expansion.  The term $\eta_A$ is a factor
which matches heavy-quark effective theory to QCD, and contains
perturbative-QCD logarithmic dependence on the heavy-quark masses; it is independent of
light quark mass.  The term proportional to $g^2_{DD^*\pi}$ is
short-hand for the one-loop staggered chiral logarithms, and is given in
the appendix for ease of reference.  The rooted staggered expression was
derived in Ref.~\cite{Laiho:2005ue}.
The one-loop staggered logarithms depend on both valence and sea quark masses,
and include taste-breaking effects coming from the light quark
sector. This expression also contains explicit dependence on the lattice
spacing $a$, and requires as inputs the parameters of the
staggered chiral lagrangian $\delta'_V$, $\delta'_A$, in addition
to the staggered taste splittings $\Delta_{P,A,T,V,I}$ \cite{Aubin:2004fs}.  These
parameters can be obtained from chiral fits to the light
pseudoscalar meson sector and are held fixed in the chiral
extrapolation of $h_{A_1}(1)$. The continuum low energy constant
$g_{DD^*\pi}$ appears, 
and below we take a generous range inspired by a 
combined fit to many different experimental inputs,
including a leading-order analysis of the $D^*$ width.
The $D^*$-$D$ splitting
$\Delta^{(c)}$ is well determined from experiment.  The only
other parameter that appears at NLO is the constant $X_A(\Lambda)$, and this
is determined by our lattice data for $h_{A_1}(1)$.

Although the lattice data are well described by the NLO
formula, it is useful to go beyond NLO and to include the next-to-next-to-leading-order (NNLO) analytic
terms as a way to estimate systematic errors.  We do not include the
NNLO logarithms because they are unknown and would require a two-loop
calculation.  The expression including analytic terms through NNLO is
\bea\label{eq:NNLO} h_{A_1}^{\textrm{NNLO}}(1)/\eta_A = 1 + \textrm{NLO} + c_1 m^2_{X_P}
+c_2\left(2m^2_{U_P}+m^2_{S_P}\right)  + c_3 a^2, \eea
\noindent where the subscript $P$ on the meson masses indicates the taste pseudo-scalar mass.  We use the notation from the rS$\chi$PT literature that $m_{X_\Xi}$ is a taste $\Xi$ meson made of two valence $x$ quarks, $m_{U_\Xi}$ is a taste $\Xi$ meson made of two light sea quarks, and $m_{S_\Xi}$ is a taste $\Xi$ meson made of two strange sea quarks.  By heavy-quark symmetry, the $c_i$ are suppressed by a factor of $1/m^2_c$.   Since the only free parameter through NLO is an overall constant, we include the NNLO analytic terms in the fit used for our central value.  This leads to a larger statistical error and is more conservative.

\section{Treatment of chiral extrapolation}

In this section, we discuss the approach we have developed to
disentangle the heavy- and light-quark discretization effects and to
perform the chiral and continuum extrapolations.  
In the Fermilab method, heavy-quark discretization errors can be 
estimated by comparing the heavy-quark expansions for lattice gauge 
theory and continuum QCD 
\cite{Kronfeld:2000ck,Harada:2001fi,Harada:2001fj,Oktay:2008ex}.
The dependence on $a$ is not simply a power series (unless $ma\ll1$), 
so power-counting estimates in HQET are used.
On the other hand, some of the light quark discretization effects are 
constrained by rS$\chi$PT.
The heavy-quark errors are asymptotically
constrained by the Symanzik low-energy Lagrangian when $m_ha \ll 1$ and
by heavy-quark symmetry even when $m_h a$ is close to 1.  In the region
in between, the errors smoothly interpolate the asymptotic behavior
\cite{ElKhadra:1996mp, Oktay:2008ex}.  The errors in the SW action used
for the heavy quarks decrease with lattice spacing as $\alpha_s a$ in
the $m_ha \ll 1$ region, as compared to the light quark (improved
staggered) discretization errors, which decrease much faster, as
$\alpha_s a^2$.

The first step of the method is to normalize the numerical data for
$h_{A_1}(1)$ to a fiducial point by forming the ratio
\bea
\label{eq:Rfid} {\cal R}_{\rm fid}(m_x, \hat{m}', m_s', a) \equiv \frac{h_{A_1}(m_x, \hat{m}', m_s',
a)}{h_{A_1}(m_x^{\rm fid}, \hat{m}^{\rm fid}, m_s^{\rm fid}, a)},
\eea
where $m^{\rm fid}$ is a fiducial mass, $m_x$ is the light (spectator)
valence quark, $\hat{m}'$ is the isospin averaged light sea quark on a
particular ensemble, and $m_s'$ is the strange sea quark on that
ensemble.  (Note that the factor of $\eta_A$ in Eqs.~(\ref{eq:NLOsChPT}) and (\ref{eq:NNLO}) cancels in the ratio.) The principle advantage of this ratio is that heavy quark
discretization effects largely cancel, since the heavy quarks are the
same in numerator and denominator.  This allows us to disentangle the
heavy-quark discretization effects from those of the light quark sector
coming from staggered chiral logarithms, thus isolating the
(taste-violating) discretization effects specific to the staggered light
quarks.  These light quark discretization effects can appear in
non-analytic terms in rS$\chi$PT and are due to violations of
taste-symmetry.  They can be removed to a given order in rS$\chi$PT (we
work to NLO) in fits to the numerical data at multiple lattice spacings
using the explicit rS$\chi$PT formula of Eq.~(\ref{eq:NNLO}), since this
formula includes the staggered lattice artifacts.  The continuum limit
of the ratio ${\cal R}_{\rm fid}$ can be obtained using our fitted
values for parameters in rS$\chi$PT and taking $a\rightarrow 0$ in the
rS$\chi$PT expression for ${\cal R}_{\rm fid}$.  We do not need a more
explicit ansatz for the functional form of the heavy-quark
discretization effects, since they largely cancel in the ratio.

Normalizing the continuum extrapolated ratio ${\cal R}_{\rm fid}$ by
$h_{A_1}$ at the fiducial point on a very fine fiducial lattice where
the heavy-quark discretization effects are small gives a value close to
the physical continuum result,
\bea\label{eq:anchorPt}
h_{A_1}(\hat{m}, \hat{m}, m_s, 0) \approx 
	h_{A_1}
	(m_x^{\rm fid}, \hat{m}^{\rm fid}, m_s^{\rm fid}, a^{\rm fid})\times
	{\cal R}_{\rm fid}(\hat{m}, \hat{m}, m_s, 0),
\eea
where the relation becomes exact as $a^{\rm fid}\rightarrow 0$.  Note
that the requirement that the heavy-quark discretization effects must be
small enforces the condition that the improved staggered light-quark
discretization effects be even smaller (and likely negligible) because
the staggered discretization effects decrease much faster with lattice
spacing.  The fiducial masses 
$m_x^{\rm fid}$, $\hat{m}^{\rm fid}$, and $m_s^{\rm fid}$
should be chosen large enough
that it would be feasible to simulate this mass point on a very fine
lattice (since the cost rises significantly as the mass of the light sea
quarks is decreased), thus normalizing the lattice data to a point where the
heavy-quark discretization effects are small.  The fiducial masses should
not be chosen so large, however, that rS$\chi$PT would not be a reliable guide
in performing the continuum and chiral extrapolation of ${\cal R}_{\rm
fid}$.  This method can be considered the crudest form of step-scaling,
but it does illustrate that one does not need lattices
which are simultaneously fine enough for $b$ quarks and large enough for
light quarks in order to simulate, with high precision, quantities that involve both.  In practice, we find $m_x^{\rm fid} = \hat{m}^{\rm fid}
\approx 0.4 m_{\rm s}$ and $m_s^{\rm fid} \approx m_{\rm s}$ are
reasonable values for the fiducial masses.
The fiducial lattice
spacing should be chosen as fine as is practical; a succession of
progressively finer fiducial lattices would be desirable for verifying
that the $a$ dependence is of the expected size.  In this work we take
our finest lattice (0.09 fm) as the fiducial lattice, but we apply
Eq.~(\ref{eq:anchorPt}) with the coarser lattices taken as fiducial
lattices in order to estimate discretization errors.  We note that the
method presented above can be applied to all calculations involving the
Fermilab treatment of heavy-quarks and staggered light quarks, not only
the $B\rightarrow D^* \ell \nu$ form factor $h_{A_1}$.  It may also be desirable to compute quantities at the fiducial point (or a succession of such points) using an even further improved action for the heavy quarks.  Once the fiducial lattice spacing is of the order 0.03-0.01 fm, even the bottom quark may be treated as a ``light'' quark with the highly improved staggered action (HISQ) \cite{Follana:2006rc} or with chiral fermions, for which mass dependent discretization effects are small.  Conserved currents could then be used for many simple heavy-light quantities, removing the need for a perturbative renormalization.

For the chiral extrapolation of $h_{A_1}$ we find
it useful to form two additional ratios,
\bea
{\cal R}_{\rm sea}(\hat{m}', m_s', a) & \equiv &
	\frac{h_{A_1}(m_x^{\rm fid}, \hat{m}', m_s', a)}{h_{A_1}(m_x^{\rm fid}, \hat{m}^{\rm fid}, m_s^{\rm fid}, a)}, 
\label{eq:Rsea}  \\
{\cal R}_{\rm val}(m_x, \hat{m}', m_s', a) & \equiv & 
	\frac{h_{A_1}(m_x, \hat{m}', m_s', a)}{h_{A_1}(m_x^{\rm fid}, \hat{m}', m_s', a)},  
\label{eq:Rval}
\eea
whose product is clearly ${\cal R}_{\rm fid}$, Eq.~(\ref{eq:Rfid}).  
${\cal R}_{\rm sea}$ and ${\cal R}_{\rm val}$ separate the sea and
valence quark mass dependence, which makes it easier to assess
systematic errors.  
The values of $h_{A_1}$
that enter Eqs.~(\ref{eq:Rsea}) and (\ref{eq:Rval}) are obtained from
\bea
\label{eq:ha1def}
h_{A_1}=\rho \sqrt{\overline{R}_{A_1}},
\eea
where $\overline{R}_{A_1}$ is the average of double ratios defined in
Eqs.~(\ref{eq:Ra1avg}).  
The ratios in Eqs.~(\ref{eq:Rsea}) and (\ref{eq:Rval}) are
now quadruple ratios, where the excited
state contamination is further suppressed over that of the double
ratio.  Performing the chiral extrapolation, taking the continuum limit of the
two ratios, and multiplying them together we recover ${\cal R}_{\rm
fid}(\hat{m},\hat{m},m_s,0)$ by construction.  Thus, we can rewrite
Eq.~(\ref{eq:anchorPt}) as
\bea
\label{eq:fidRseaRval}
h_{A_1}^{\rm phys}\approx h_{A_1}(m_x^{\rm fid}, \hat{m}^{\rm fid}, m_s^{\rm fid}, a^{\rm fid})\times [{\cal R}_{\rm sea}(\hat{m},
m_s, 0)\times {\cal R}_{\rm val}(\hat{m}, \hat{m}, m_s, 0)],
\eea
where, again, the relation becomes exact as $a^{\rm fid}\rightarrow 0$.

To the extent that the extrapolation in sea quark masses is mild, the
ratio ${\cal R}_{\rm sea}$ should be close to one, since the valence light
mass is the same in both numerator and denominator. ${\cal R}_{\rm val}$
contains less trivial chiral behavior.  However, since the
numerator and denominator are computed on the same ensemble (with
different valence masses), they are correlated, and statistical errors tend to cancel in
${\cal R}_{\rm val}$.  The ratio
${\cal R}_{\rm sea}$ has small statistical errors because the 
valence mass~$m_x^{\rm fid}$ in
that ratio is relatively heavy.  Of course, the heavy-quark
discretization errors are significantly suppressed in both ratios,
isolating the light quark mass dependence and staggered
discretization effects.  A direct chiral fit to the numerical data (not
involving the ratios introduced here) would require a more explicit
ansatz for the treatment of the heavy quark discretization effects than
is needed in the ratio fits\footnote{  A direct (correlated) chiral fit would
still, however, reflect the correlations which cause cancellations in
the statistical errors in the ratios.}. Note that in the ratios the
fiducial point need not be tuned to the same mass at every lattice
spacing; differences can be accounted for in the fit itself.  
The fiducial points used at different lattice spacings are 
$m_x^{\rm fid} = \hat{m}^{\rm fid} = 0.4 m'_{\rm s}$ and 
$m_s^{\rm fid} = m'_{\rm s}$.  The explicit values are given in 
Table~\ref{tab:fid}, along with the calculated values of 
$\sqrt{\overline{R}_{A_1}}$ and $h_{A_1}^{\rm fid}$ at that fiducial point.
\begin{table}
\begin{center}
\caption{Fiducial masses used at the three different lattice spacings.
The first four columns are the approximate lattice spacing in fm, the
fiducial valence quark mass, the fiducial light sea quark mass, and the
fiducial strange quark mass.  The fifth and sixth columns are the values of
$\sqrt{\overline{R}_{A_1}}$ and $h_{A_1}^{\rm fid}$, respectively, computed at that fiducial point.}
 \label{tab:fid}
\begin{tabular}{cccccc}
\hline \hline
 lattice spacing (fm) & $am_x^{\rm fid}$ & $a\hat{m}^{\rm fid}$ & $am_s^{\rm fid}$ & $\sqrt{\overline{R}_{A_1}}$ & $h_{A_1}^{\rm fid}$ \\
  \hline
  $0.15$  &  0.0194 \ \ & 0.0194 \ \ & 0.0484 & 0.9211(73) & 0.9180(73) \\
  $0.12$ & 0.02 & 0.02 & 0.05  & 0.9112(73) & 0.9079(73) \\
  $0.09$ & 0.0124 & 0.0124 & 0.031  & 0.9210(85) & 0.9237(85) \\
\hline\hline
\end{tabular}
\end{center}
\end{table}

The constant term $X_A(\Lambda_\chi)$ in Eq.~(\ref{eq:NLOsChPT}) cancels
in the ratios ${\cal R}_{\rm sea}$ and ${\cal R}_{\rm val}$, so the
behavior of these ratios is completely predicted through NLO in the
chiral expansion.  We find good agreement between the predicted form and
the numerical data.  However, given that our fiducial spectator quark
mass is rather large (around $0.4 m_{s}$), we include the NNLO analytic
terms in the ratio fits in order to estimate systematic errors
associated with the chiral expansion.  There are only two
new continuum low energy constants introduced at this higher order, and
the ratios ${\cal R}_{\rm sea}$ and ${\cal R}_{\rm val}$ determine one
each.  There is also an analytic term proportional to $a^2$ appearing at
this order, but it cancels in each of the ${\cal R}_{\rm sea}$ and ${\cal
R}_{\rm val}$ ratios.

In future calculations, it would be feasible to use a much finer lattice
spacing for the fiducial point, 
thereby further reducing
heavy-quark discretization errors.  
For now, however, we use  
$h_{A_1}(m_x^{\rm fid}, \hat{m}^{\rm fid}, m_s^{\rm fid}, 0.09~{\rm fm})$,
with the fiducial masses in Table~\ref{tab:fid}, 
in Eq.~(\ref{eq:fidRseaRval}).
As a way to estimate discretization errors we use our results for
$h^{\rm fid}_{A_1}$ at the two coarser lattice spacings in
Eq.~(\ref{eq:fidRseaRval}) also.

At the lattice 
spacings used in this work the
light-quark discretization effects may still be non-negligible compared to
heavy-quark discretization effects.
With rS$\chi$PT it is possible to remove from $h^{\rm fid}_{A_1}$ 
the discretization effects associated with staggered chiral
logarithms, although purely analytic discretization errors remain.
Removing this subset of staggered effects leads to a value for the
fiducial form-factor which we call the ``taste-violations-out'' value.
Not removing them leads to the ``taste-violations-in''
value.  The difference turns out to be negligible, less than $0.1\%$ on
our coarsest ensemble and less than $0.01\%$ on the fine ensemble.
Thus, the discretization effects in our lattice data coming from
taste-violations in the staggered chiral logarithms are extremely small
at the fiducial point mass, and we neglect this difference in the analysis.

\begin{figure}
\begin{center}
\includegraphics[scale=.55]{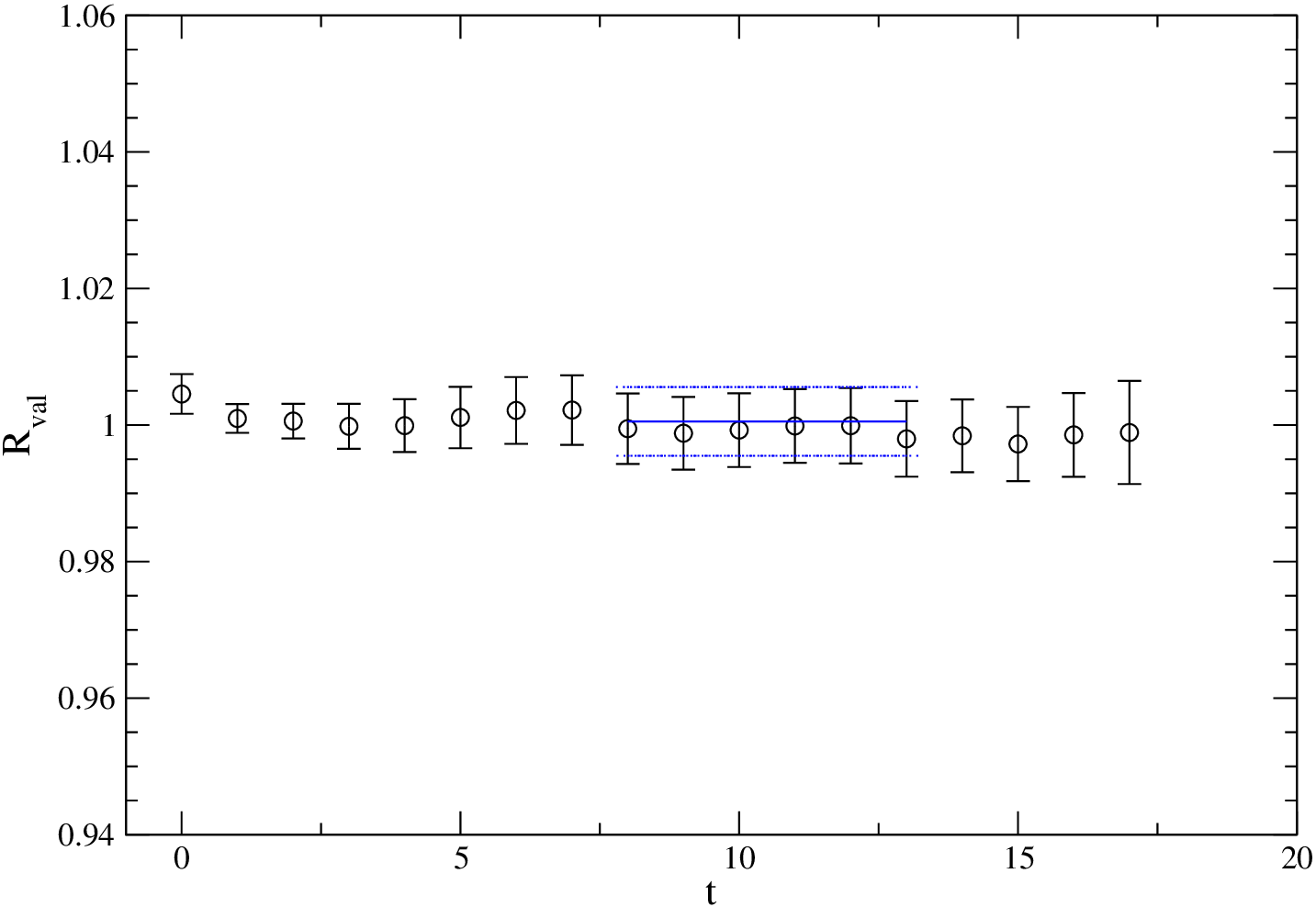}
\caption{${\cal R}_{\rm val}$ on the $a\hat{m}'= 0.0062$ fine ensemble.
The valence mass in the numerator is the full QCD value of $am'_x=
0.0062$ while the fiducial valence mass in the denominator is $am^{\rm
fid}_x=0.0124$.  The fit to a constant has a $\chi^2/\rm{d.o.f} =0.20$.}
\label{fig:Rval}
\end{center}
\end{figure}

Figure~\ref{fig:Rval} shows the plateau fit to the ratio 
${\cal R}_{\rm val}$ on 
the fine ensemble with $(a\hat{m}',am'_s)=(0.0062,0.031)$.
\begin{table}
\begin{center}
\caption{Computed values of ${\cal R}_{\rm sea}$.  The first three
columns are the arguments of ${\cal R}_{\rm sea}$ as defined in Eq.~(\ref{eq:Rsea}); they are
the light sea quark mass $\hat{m}'$, the strange quark mass $m_s'$, and the approximate lattice spacing in fm.
The fourth column is ${\cal R}_{\rm sea}$.}
 \label{tab:Rsea}
\begin{tabular}{cccc}
\hline \hline
  $a\hat{m}'$ & $am_s'$ & $a$(fm) & ${\cal R}_{\rm sea}$ \\
  \hline
   0.0097 \ \ & 0.0484 & $0.15$  & 1.009(12)\\
    0.01 & 0.05  & $0.12$ & 1.0070(98)\\
     0.007 & 0.05  & $0.12$ & 1.0027(91)\\
  0.005 & 0.05  & $0.12$ & 1.014(10)\\
   0.0062 & 0.031  & $0.09$ & 1.000(12) \\
    0.0031 & 0.031  & $0.09$ & 0.996(10) \\
\hline\hline
\end{tabular}
\end{center}
\end{table}
The valence mass in the
numerator is the full QCD value of $am'_x= 0.0062$, while the fiducial
valence mass in the denominator is $am^{\rm fid}_x=0.0124$.  Both
numerator and denominator are computed on the same ensemble, so they
have the same sea quark masses, and correlated statistical errors
largely cancel in the ratio, as expected.  Excited-state contamination
is also reduced.  Computed values for ${\cal R}_{\rm sea}$ on all of our
ensembles are given in Table~\ref{tab:Rsea}, 
\begin{table}
\begin{center}
\caption{Computed values of ${\cal R}_{\rm val}$.  The first four
columns are the arguments of ${\cal R}_{\rm val}$ as defined in Eq.~(\ref{eq:Rval}); they are
the light valence quark mass $m_x$, the light sea quark mass $\hat{m}'$,
the strange quark mass $m_s'$, and the approximate lattice spacing in fm.  The fifth column is ${\cal R}_{\rm
val}$.}
 \label{tab:Rval}
\begin{tabular}{ccccc}
\hline \hline
  $am_x$ & $a\hat{m}'$ & $am_s'$ & $a$(fm) & ${\cal R}_{\rm val}$ \\
  \hline
    0.0097 \ \ & 0.0097 \ \ & 0.0484 & $0.15$  & 1.0056(65)\\
   0.01 & 0.01 & 0.05  & $0.12$ & 0.9994(41)\\
    0.007 & 0.007 & 0.05  & $0.12$ & 0.9900(57)\\
     0.005 & 0.005 & 0.05  & $0.12$ & 1.0081(90)\\
   0.0062 & 0.0062 & 0.031  & $0.09$ & 1.0005(50) \\
   0.0031 & 0.0031 & 0.031  & $0.09$ &  1.0043(62) \\
\hline\hline
\end{tabular}
\end{center}
\end{table}
and the computed values for ${\cal R}_{\rm val}$ are given in
Table~\ref{tab:Rval}.

\section{Systematic errors}

In the following subsections, we examine the uncertainties in our
calculation due to fitting and excited states, the heavy-quark
mass dependence, the chiral extrapolation of the light spectator
quark mass, discretization errors, and perturbation theory.  As
mentioned in Section II, statistical uncertainties are computed
with a single elimination jackknife and the full covariance
matrix.

\subsection{Fitting and excited states}

We have examined plateau fits to the time dependence of the double and
quadruple ratios introduced in Sections I and V. The $\chi^2$ in our
fits is defined with the full covariance matrix.
The fits to the ratios were done under a
single elimination jackknife, after blocking the numerical data by 8 on the fine
lattices and by 4 on the coarse and coarser lattices.  The blocking
procedure averages 4 (or 8) successive configurations before performing
the single elimination jackknife.  These values for the block size were
chosen such that the statistical error on the double ratio fit did not
increase when a larger block size was used.  Statistical errors were
determined in fits that included the full correlation matrix, which was remade for each jackknife fit.  The jackknife data sets on different ensembles were then combined into a
larger block-diagonal jackknife data set in order to perform the chiral
fits.  In this way, the fully correlated statistical errors were
propagated through to the final result.

With our high statistics (several hundred lattice gauge field
configurations for each ensemble), we are
able to resolve the full covariance matrix well enough that we do not
need to apply a singular value decomposition cut on the eigenvalues of
the covariance matrix.  The double ratio fit is needed to establish
$h_{A_1}(1)$ at the fiducial point (which was computed on the 0.0124/0.031
fine ensemble), while the quadruple ratios, ${\cal R}_{\rm val}$ and
${\cal R}_{\rm sea}$ are computed on the other ensembles in order to
perform the chiral extrapolation and to remove taste breaking
non-analytic terms.  We find that the fit to the double ratio at the
fiducial point on the 0.0124/0.031 ensemble is well described by a constant
over a range of seven time slices.  The excited state contamination in
the quadruple ratios is even further suppressed, and we find that the
correlated $\chi^2$ values allow for a constant fit region of six to
ten time slices, depending upon the lattice spacing.  We take the good
correlated $\chi^2/\textrm{d.o.f.}$, ranging from 0.15 to 1.00, in our
constant plateau fits as evidence that the excited state contamination
in these fits is negligible as compared to other errors.

As an additional check of the jackknife fitting procedure, bootstrap
fits were done to all of the double and quadruple ratios needed for this
work.  Close agreement was found for both central values and statistical
errors.  The statistical errors were typically the same size within $10\%$, and central values were well within 1$\sigma$.  The jackknife procedure had slightly larger errors than that of the bootstrap.

\subsection{Heavy-quark mass dependence}

\begin{table}
\begin{center}
\caption{Errors in the $\kappa_{b,c}$ parameters.  The first column labels the heavy quark, the second gives the statistical and fitting error for the $\kappa$ parameter, the third gives the discretization error, and the fourth combines these in quadrature.}
\label{tab:kappa_errors}
\begin{tabular}{cccc}
\hline \hline
  $\kappa$ & \ \ \  statistics + fitting \ \ \ & discretization \ \ & total \\
  \hline
  $\kappa_c$ &  $1.2\%$ & $0.3\%$ & $1.2\%$  \\
  $\kappa_b$ &  $5.6\%$  & $1.3\%$ & $5.7\%$ \\
 \hline \hline
\end{tabular}
\end{center}
\end{table}

The value for $h_{A_1}(1)$ depends on the heavy-quark masses, which are
set by tuning the hopping parameters $\kappa_b$ and $\kappa_c$.  The
principal method starts by fitting the lattice pole energy to $E(\mathbf{p})$ to the dispersion relation, 
\bea
	E(\mathbf{p})=M_1 + \frac{\mathbf{p}^2}{2M_2} +
		b_1\mathbf{p}^4+b_2\sum^{3}_{j=1}|p_j|^4 + \cdots, 
\eea 
in order to obtain the kinetic mass $M_2$ (as well as $b_1$ and $b_2$, which
are unimportant here).  In the Fermilab method \cite{ElKhadra:1996mp,Kronfeld:2000ck,Harada:2001fj}, $\kappa$ is adjusted so
that the kinetic mass agrees with experiment.  Here we take the
spin-average of kinetic masses of pseudoscalar and vector heavy-strange
mesons and obtain our central values for $\kappa_b$ or $\kappa_c$, 
respectively, from the (spin-averaged) $B^{(*)}_s$ and $D^{(*)}_s$
masses.  Applying this procedure we find statistical and fitting errors of $5.6\%$ for $\kappa_b$ and $1.2\%$ for $\kappa_c$ on the fine ($a=0.09$ fm) ensembles.  There is an additional error in $\kappa$ due to discretization effects.  We determine this error by estimating the size of discretization effects for the Fermilab action (at $a=0.09$ fm) as in Ref.~\cite{Kronfeld:1996uy}.  This error is $1.3\%$ for $\kappa_b$ and $0.3\%$ for $\kappa_c$.  Adding in quadrature the statistical and fitting error together with the discretization error leads to a total relative uncertainty of $5.7\%$ for $\kappa_b$ and $1.2\%$ for $\kappa_c$.  This error budget is summarized in Table~\ref{tab:kappa_errors}.  Note that these errors are conservative and are likely to decrease substantially with more sophisticated fitting methods and the higher statistics data set currently being generated.

\begin{figure}
\begin{center}
	\includegraphics[scale=.55]{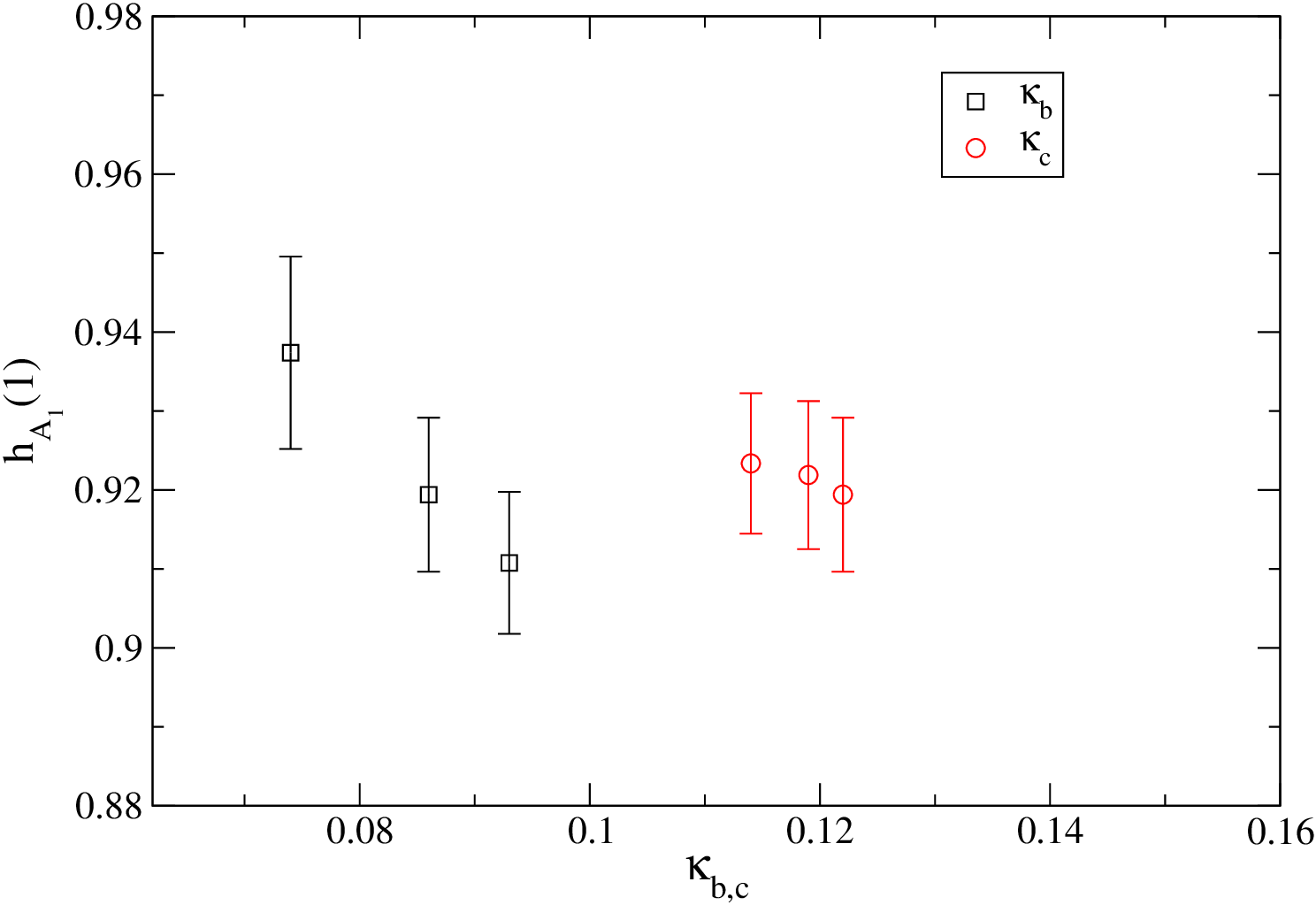}
	\caption{$h_{A_1}(1)$ for different $\kappa_{h}$ values on the
	coarse $\hat{m}'=0.02$ ensemble (full QCD point).  The points
	labelled $\kappa_b$ show how $h_{A_1}(1)$ depends on $\kappa_b$ when $\kappa_c$ is
	fixed to its tuned value. For the points labelled $\kappa_c$ the 
	roles of $\kappa_b$ and $\kappa_c$ are reversed.}
	\label{fig:kappa}
\end{center}
\end{figure}

We have computed $h_{A_1}(1)$ at several different values of the bare
charm and bottom quark masses, and these simulated points can be used to
estimate the error in $h_{A_1}(1)$ from the above uncertainties in the
tuning of the heavy-quark $\kappa$ values.  Figure~\ref{fig:kappa}
illustrates the dependence of $h_{A_1}(1)$ as a function of bottom and
charm quark $\kappa$ values on one of the coarse ($a=0.12$ fm)
ensembles.
The points labelled $\kappa_b$ show $h_{A_1}(1)$ where we have fixed
$\kappa_c$ to the tuned charm value, but vary the bare $\kappa_b$ along
the $x$-axis.  The points labelled $\kappa_c$ are similar, where the value
of $\kappa_b$ is fixed at its tuned value, and the bare $\kappa_c$ is
varied.  The above uncertainties in the $\kappa$'s, combined with the
variation of $h_{A_1}(1)$ with $\kappa$, lead to a systematic error of
$0.7\%$ in $h_{A_1}(1)$, labelled ``kappa tuning'' in
Table~\ref{tab:errors}.

\subsection{Perturbation theory}

The perturbative calculation of $\rho_{A_1}$ is needed to
match the heavy-quark lattice current, and the
calculation has been carried out to one-loop order [$O(\alpha_s)$].  As
discussed in Section IV, much of the renormalization cancels when
forming the ratios of $Z$ factors that define $\rho$
[Eq.~(\ref{eq:rho})], and the coefficients of the perturbation series
are small, by construction.  The one-loop correction is quite small,
only $0.3$--$0.4\%$ on the different lattice spacings.  We take the
entire one-loop correction of $0.3\%$ on the fine lattices as an
estimate of the error introduced by neglecting higher orders in the
perturbative expansion.

\subsection{Chiral extrapolation}

We estimate our systematic error due to the chiral extrapolation by comparing fits with and without the additional terms with coefficients $c_i$ in Eq.~(\ref{eq:NNLO}), i.e.~analytic terms of higher order than NLO in rS$\chi$PT, since the
two-loop NNLO logarithms are unknown.  We also compare with continuum $\chi$PT, both NLO and (partial) NNLO.  There are additional errors due to the uncertainties in the parameters that enter the NLO rS$\chi$PT formulas.  By far the largest uncertainty of this kind is that due to the uncertainty in $g_{DD^*\pi}$. Finally, there is an error due to a mistuning of the parameter $u_0$ on the coarse lattices.  All of these errors are discussed below in more detail.  In the discussion of chiral extrapolation errors, it is important to keep in mind that the chiral logarithms (either rS$\chi$PT or continuum) are tiny ($\sim 3\times 10^{-3}$) in the region where we have data.  Non-analytic behavior is important only near the physical pion mass where the $\chi$PT should be a very good description in the continuum.  The main feature of the chiral extrapolation is a cusp that appears close to the physical pion mass (in the valence sector),
due to the $D\pi$ threshold and the fact that the $D$-$D^*$ splitting is
very close to the physical pion mass.  This cusp represents real physics, and must be included in any version of the chiral extrapolation used to estimate systematic errors.

We extrapolate the light sea and light valence quark
masses from the values used in the simulations, between $m_s/2$ and
$m_s/10$, to the average physical light quark mass, around $m_s/27$.  We
use staggered chiral perturbation theory and the
prescription introduced in Section VI to remove the non-analytic
taste-breaking discretization effects coming from the staggered light
quark sector.  Separate fits are performed for the two ratios introduced
in Eqs.~(\ref{eq:Rsea}) and (\ref{eq:Rval}), ${\cal R}_{\rm sea}$ and
${\cal R}_{\rm val}$.  The chiral extrapolation is performed on these
ratios, and the staggered discretization errors appearing in the NLO
chiral logarithms are removed by taking $a\rightarrow 0$ in the
rS$\chi$PT expression.  With the NNLO analytic terms given in
Eq.~(\ref{eq:NNLO}) the chiral extrapolation formulas for the ratios
are
\bea
\label{eq:RvalNNLO} && {\cal R}_{\rm val} = 1 + \textrm{NLO}_{\rm logs} + c_1 m^2_{X_P},  \\ &&
\label{eq:RseaNNLO} {\cal R}_{\rm sea}= 1 + \textrm{NLO}_{\rm logs} + c_2 (2m^2_{U_P}+m^2_{S_P}),
\eea
where 
$\textrm{NLO}_{\rm logs}$ is a schematic notation representing
the chiral logarithms coming from numerator and denominator.  These
terms are different for the two ratios, and can be obtained
straightforwardly from the definitions of the ratios
Eqs.~(\ref{eq:Rsea}) and (\ref{eq:Rval}), and the formula for the
non-analytic terms in Eq~(\ref{eq:schpt}).  The formula for ${\cal R}_{\rm val}$ in the continuum is given explicitly in Eq.~(\ref{eq:chptRval}), for the purposes of illustration.  The NNLO term $c_3 a^2$ in
Eq.~(\ref{eq:NNLO}) cancels in the ratios, and ${\cal R}_{\rm sea}$ and
${\cal R}_{\rm val}$ each determine one of the remaining two NNLO
coefficients.  Note that the factor of $\eta_A$ in Eqs.~(\ref{eq:NLOsChPT}) and (\ref{eq:NNLO}) cancels in the chiral formulas for the two ratios.  The only free parameters in our chiral fits are $c_1$ and $c_2$; the rest are determined from phenomenology or from rS$\chi$PT fits to the pseudoscalar sector.

\begin{figure}
\begin{center}
\includegraphics[scale=.55]{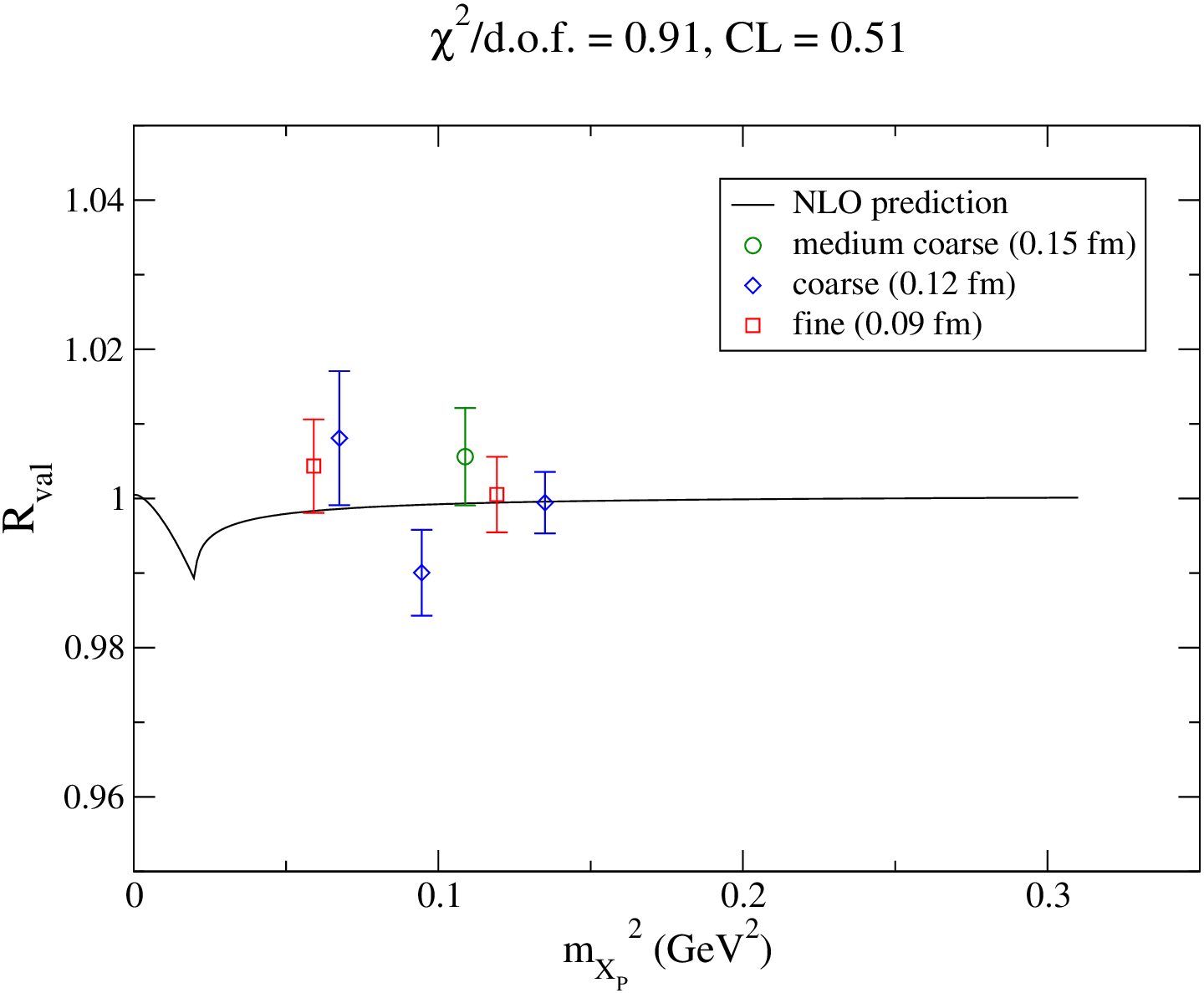}
\caption{${\cal R}_{\rm val}$ ratio versus valence pion mass squared on all ensembles for the three different lattice spacings.  The curve is the continuum prediction through NLO in continuum $\chi$PT for this quantity.  (See Appendix.) \label{fig:chiralRvalNLO}}
\end{center}
\end{figure}

\begin{figure}
\begin{center}
\includegraphics[scale=.55]{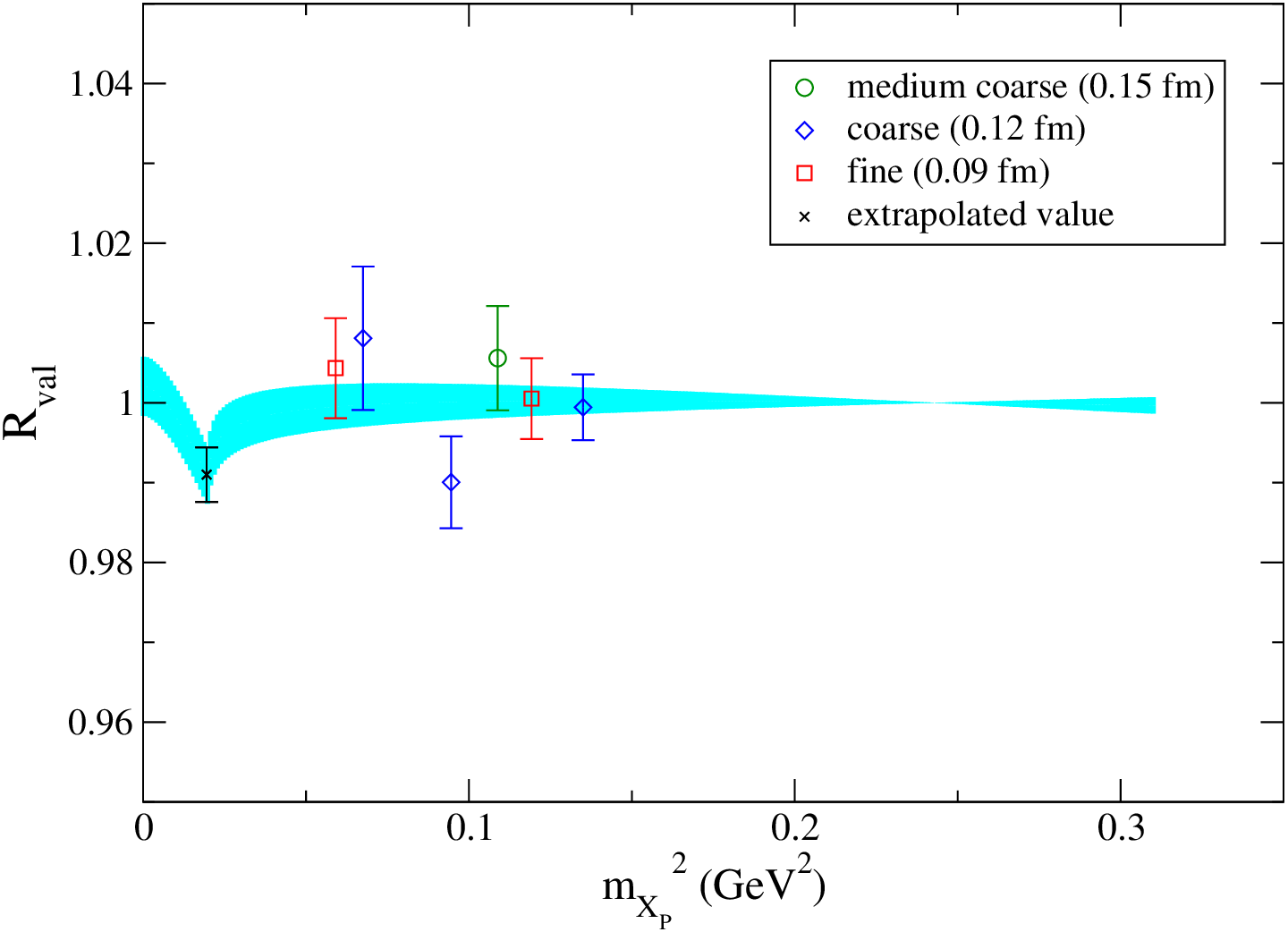}
\caption{${\cal R}_{\rm val}$ ratio versus valence pion mass squared on all ensembles for the three different lattice spacings.  The curve is the fit with 1 sigma error band to
the ratio for all three lattice spacings using rS$\chi$PT,
extrapolated to the continuum by taking $a\rightarrow 0$ in the NLO staggered chiral logarithms.   \label{fig:chiralRval}}
\end{center}
\end{figure}

  The ratios in Eqs.~(\ref{eq:Rsea}) and (\ref{eq:Rval}) are completely predicted through NLO in the continuum once $f_\pi$, $g_{DD^*\pi}$, and the $D$-$D^*$ splitting $\Delta^{(c)}$ are taken from experiment.  The constants $f_\pi$ and $g_{DD^*\pi}$ appear in an overall multiplicative factor $\frac{g^2_{DD^*\pi}}{48\pi^2f_\pi^2}$ in front of the logarithmic term, as can be seen in Eq.~(\ref{eq:schpt}) and Eq.~(\ref{eq:chptRval}).  We take a fairly conservative range for the constant $g_{DD^*\pi}$ determined from phenomenology, as discussed below, and the errors in this quantity are accounted for in our final error budget.  In the mass region where we have data, the NLO continuum chiral logarithms contribute to $h_{A_1}(1)$ at the $\sim 3\times 10^{-3}$ level or less.  Figure~\ref{fig:chiralRvalNLO} illustrates this, where the NLO continuum $\chi$PT prediction Eq.~(\ref{eq:chptRval}) is plotted over our data points for ${\cal R}_{\rm val}$.  We find that the NLO continuum $\chi$PT describes the data quite well, giving a $\chi^2/\textrm{d.o.f.}=0.91$ and a corresponding CL=0.51.  This result is unchanged in the rS$\chi$PT fits; the effects of staggering are negligible in the region where we have data.  We include the term proportional to $c_1$ in Eq.~(\ref{eq:RvalNNLO}) in our fits used to obtain the central value for this quantity, as explained in Section~V.  (Since including a linear term proportional to $c_1$ increases the statistical error in $h_{A_1}$, we take our central value and statistical error from this fit to be conservative.)  This ``partial NNLO'' fit also has a good $\chi^2/\textrm{d.o.f.}=1.05$, with a corresponding CL=0.39.  The constant linear term is small and consistent with zero [$c_1=-0.006(15)$].  Figure~\ref{fig:chiralRval} shows the fit to ${\cal R}_{\rm val}$ versus
$m^2_{X_P}$ for all three lattice spacings using the rS$\chi$PT formula,
Eq.~(\ref{eq:RvalNNLO}).

Although the data for ${\cal R}_{\rm val}$ is consistent with a constant, the cusp appearing close to the physical pion mass is a prediction of NLO $\chi$PT and has a physical origin, namely the $D$-$\pi$ threshold, as we have remarked.  Thus, any fits used to estimate systematic errors, even those that are somewhat ad hoc, such as those including higher order polynomial terms, must include this cusp.  Note that the cusp appears at the physical pion mass (in either SU(3) or SU(2) $\chi$PT), and is therefore in a region where $\chi$PT is expected to be a reliable expansion.  The cusp is a property of the function $F(m, \Delta^{(c)}/m)$ given in Eq.~(\ref{eq:F}), and the position of the cusp as a function of $m_{X_P}^2$ is determined by the $D$-$D^*$ splitting $\Delta^{(c)}$ and the physical pion mass.  We take these two quantities from experiment rather than from the lattice, since the experimental uncertainties are much smaller.


We find that with or without the
NNLO analytic terms, the $\chi$PT (continuum or rooted staggered) describes the lattice data with $\chi^2/\textrm{d.o.f.}$ close to 1 and correspondingly good confidence levels.   We find a confidence level for the fit to ${\cal R}_{\rm sea}$ of 0.76 for
the fit that includes NNLO analytic terms. The strictly NLO expression for the lattice ratio ${\cal R}_{\rm sea}$ has no free parameters, but it describes the data with a confidence level of 0.73.   Similar fits to ${\cal R}_{\rm val}$ are described above and yield reasonable confidence levels for both types of fits.  Since the lattice data do not distinguish
between these model fit functions, and the fit using only the NNLO
analytic terms is not systematic in the chiral expansion, we assign the
difference between the two determinations, which is $0.9\%$, as the systematic error of
leaving out higher order terms when performing the chiral
extrapolation.  The final results for ${\cal R}_{\rm sea}$,
${\cal R}_{\rm val}$, and ${\cal R}_{\rm fid}$ are given in
Table~\ref{tab:Rs}.  The errors are statistical only; note that the strictly
NLO values have no free parameters, and therefore no statistical errors.  The final value of $h_{A_1}$ still has statistical errors coming from the statistical errors in $h_{A_1}^{\rm fid}$.  The extrapolated results for ${\cal R}_{\rm fid}$ are consistent within the statistical errors of the NNLO fit.  Again, we choose for our central value the result from the NNLO extrapolation with its larger errors to be conservative.

\begin{table}
\begin{center}
\caption{Continuum extrapolated values of ${\cal R}_{\rm sea}$, ${\cal
R}_{\rm val}$, ${\cal R}_{\rm fid}$, and $h_{A_1}(1)$ evaluated at the physical quark
masses.  The first column labels the quantity.  The second is the computed
value including NNLO analytic terms in the chiral fit.
The third is the quantity evaluated in purely NLO $\chi$PT, and has no free
parameters (once $g_{DD^*\pi}$, $f_\pi$ and $\Delta^{(c)}$ are taken from
phenomenology) in the chiral fit.  The final row shows $h_{A_1}(1)$, which includes a statistical error coming from $h_{A_1}^{\rm fid}$.  The numbers are the same to the quoted precision using rS$\chi$PT or continuum $\chi$PT.}
\label{tab:Rs}
\begin{tabular}{ccc}
\hline \hline
 \ \ & w/ NNLO \ \ &  strictly NLO  \\
  \hline
  ${\cal R}_{\rm sea}$\ \  & 1.0059(90)\ \ & 0.9983 \\
   ${\cal R}_{\rm val}$ & 0.9910(34) & 0.9895  \\
    ${\cal R_{\rm fid}}$ & 0.997(10) & 0.9878 \\
    $h_{A_1}(1)$ & 0.921(13) & 0.9124(84) \\
\hline\hline
\end{tabular}
\end{center}
\end{table}

The cyan (gray) band in Figure~\ref{fig:chiralRval} is
the continuum extrapolation with $a\to 0$ in the rS$\chi$PT
formula.  For this quantity, the staggered lattice artifacts affecting
the chiral logarithms in $h_{A_1}$ are negligible in the region where we have lattice data,
which is due mainly to the small size of the chiral logarithms themselves.  This is confirmed by the close agreement between the data
points at each lattice spacing and the continuum curve.  In fact, if we
use continuum $\chi$PT to perform the chiral extrapolation, the result
is unchanged.  The primary difference between the rS$\chi$PT expression
and the continuum $\chi$PT expression is the reduction of the cusp near
the physical pion mass in rS$\chi$PT, though our lattice data are not near enough to the
physical pion mass to demonstrate this effect.  

\begin{figure}
\begin{center}
\includegraphics[scale=.55]{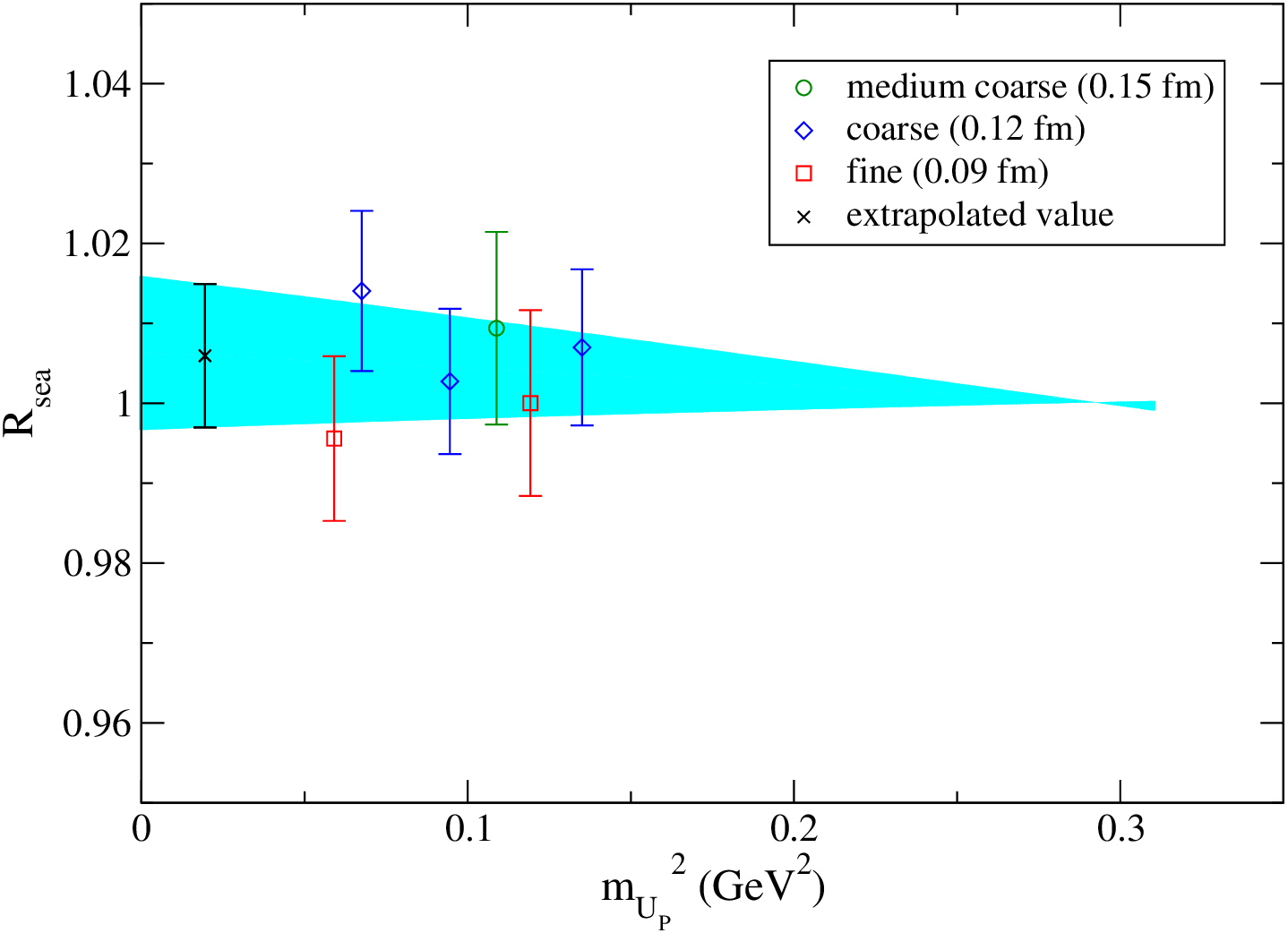}
\caption{${\cal R}_{\rm sea}$ ratio versus $m^2_{U_P}$ for all ensembles and lattice spacings.
The curve is the fit to all of the lattice data, extrapolated to the
continuum.  The curve is also extrapolated to the physical strange
sea quark mass.   \label{fig:chiralRsea}}
\end{center}
\end{figure}

Figure~\ref{fig:chiralRsea}
shows the fit to ${\cal R}_{\rm sea}$, extrapolated to the continuum and
to the physical strange sea quark mass.  Note that this ratio does not
produce a cancellation of correlations between numerator and
denominator and so has larger statistical errors than ${\cal R}_{\rm
val}$.  Here again the discretization effects due to staggered
logarithms are negligibly small.  Since the effects of including
staggered discretization effects in the chiral logarithms are negligible
in the region where we have numerical data, and since the only
nontrivial feature in the chiral extrapolation is the cusp near the
physical pion mass, which we describe by continuum $\chi$PT (our
extrapolated curve has $a\to 0$ in the rS$\chi$PT formula and thus
reduces to the continuum form), we conclude that staggered
taste-violating effects appearing in chiral logarithms are essentially
removed in our ratio extrapolations.

\begin{figure}
\begin{center}
\includegraphics[scale=.55]{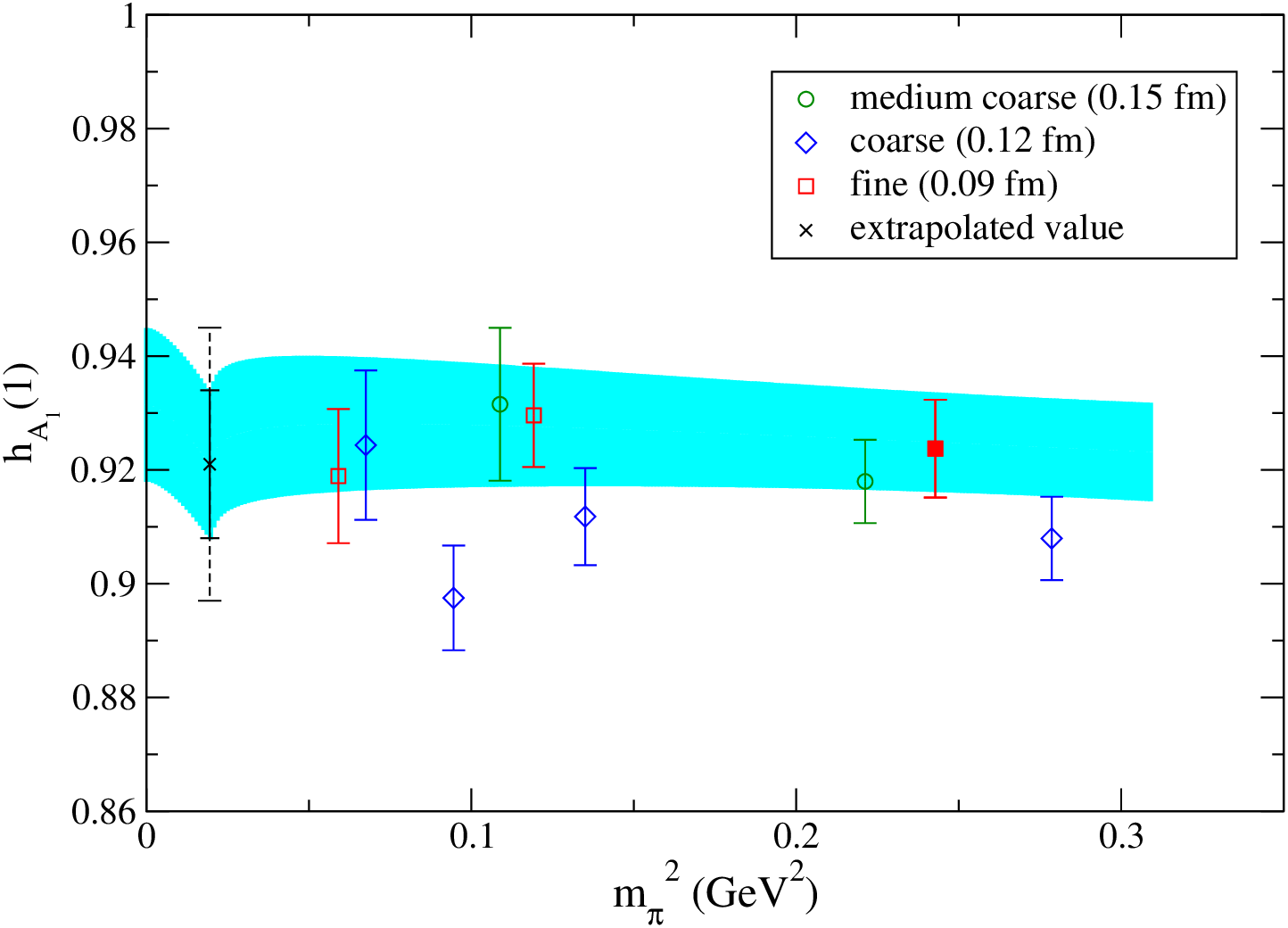}
\caption{The full QCD points versus $m^2_{\pi}$ on the three lattice spacings are shown in comparison to the continuum curve.
The curve is the product of the two continuum extrapolated ratio
fits shown in Figs.~(\ref{fig:chiralRval}) and
(\ref{fig:chiralRsea}), times the fiducial point, which we have chosen to
be the $\hat{m}'=0.0124$ fine lattice point (the filled square).  The curve is already extrapolated to the physical strange sea quark mass, and so does not perfectly overlap with the
lattice data point at the fiducial value.  The cross is the extrapolated value,
where the solid line is the statistical error, and the dashed line
is the total systematic error added to the statistical error in
quadrature. \label{fig:chiralFullQCD}}
\end{center}
\end{figure}

Figure~\ref{fig:chiralFullQCD} shows all of the full QCD points on the three lattice spacings.
The curve is the quantity,
\bea\label{eq:fid} h^{\rm phys}_{A_1}(\hat{m}') \approx h_{A_1}^{\rm fid}(m_x^{\rm fid}, \hat{m}^{\rm fid}, m_s^{\rm fid}, a^{\rm fid})\times [{\cal R}_{\rm sea}(\hat{m}',
m_s, 0)\times {\cal R}_{\rm val}(\hat{m}', \hat{m}', m_s, 0)], 
\eea
\noindent which again becomes an exact relation for the physical form factor when $a^{\rm fid}\to 0$.  The curve is thus the product of the two continuum extrapolated ratio fits shown in Figures~\ref{fig:chiralRval} and \ref{fig:chiralRsea}, times the fiducial point, which we take to
be $a\hat{m}^{\rm fid}=0.0124$ at the fine lattice spacing (the solid square in Figure~\ref{fig:chiralFullQCD}).  Because this is a full QCD curve, the valence mass $m_x$ equals the light sea mass $\hat{m}'$.  The other full QCD points are shown as open symbols in Figure~\ref{fig:chiralFullQCD} for comparison, though the fits were performed on the ratios and normalized by the fiducial point at $a\hat{m}^{\rm fid}=0.0124$.  Note that the curve is already extrapolated in the strange sea quark mass, and so does not perfectly overlap with the
$a\hat{m}^{\rm fid}=0.0124$ point.  As discussed above, when this quantity is evaluated at $\hat{m}'=\hat{m}$ it yields the value of $h_{A_1}$ at physical quark masses.  
The cross is the extrapolated value,
where the solid line is the statistical error, and the dashed line
is the total systematic error added to the statistical error in
quadrature.

The low energy constant $g_{DD^*\pi}$ enters the chiral extrapolation
formula and determines the size of the cusp near the physical pion mass.
Our data do not constrain this constant, so we
take a wide range for $g_{DD^*\pi}$ that encompasses the range of values
coming from phenomenology and lattice calculations: fits to a wide range
of experimental data prior to the measurement of the $D^*$ width by
Stewart ($g_{DD^*\pi}=0.27^{+0.06}_{-0.03}$ \cite{Stewart:1998ke}), an
update of the Stewart analysis including the $D^*$ width
($g_{DD^*\pi}=0.51$; no error quoted \cite{Arnesen:2005ez}), quark
models ($g_{DD^*\pi}\approx 0.38$ \cite{Casalbuoni:1996pg}), quenched
lattice QCD ($g^{N_f=0}_{DD^*\pi}=0.67\pm 0.08^{+0.04}_{-0.06}$
\cite{Abada:2002vj}), two flavor lattice QCD in the static limit
($g^{N_f=2}_{\rm static}=0.516\pm 0.051$ \cite{Ohki:2008py}), and the
measurement of the $D^*$ width ($g_{DD^*\pi}=0.59\pm 0.07$
\cite{Anastassov:2001cw}).  There are as of yet no 2+1 flavor lattice
calculations of $g_{DD^*\pi}$.
For this work we take $g_{DD^*\pi}=0.51\pm 0.2$, leading to a
parametric uncertainty of $0.9\%$ in $h_{A_1}(1)$ that is included as
a systematic error.

The additional low energy constants that enter the chiral formulas are
the tree-level continuum coefficients $\mu_0$ and $f$, and the
taste-violating parameters that vanish in the continuum.  These are the
taste splittings, $a^2\Delta_\Xi$ with $\Xi=P,A,T,V,I$, and the
taste-violating hairpin-coefficients, $a^2\delta'_A$ and $a^2\delta'_V$.
We set $f$ to the experimental value of the pion decay constant,
$f_\pi=0.1307$ GeV, in the coefficient of the NLO logarithms.  The pion
masses used as inputs in the rS$\chi$PT formulas are computed from the
bare quark masses and converted into physical units using
\bea 
m^2_{xy}=(r_1/r^{\rm phys}_1)^2 \mu_{\rm tree}(m_x +m_y),
\eea
where $\mu_{\rm tree}$ is obtained from fits to the light pseudo-scalar
mass squared to the tree-level form (in $r_1$ units), 
$r_1^2\mu_{\rm tree}(m_x+m_y)$.  This accounts for higher-order chiral corrections and
is more accurate than using $\mu$ obtained in the chiral limit, giving a
better approximation to the pion mass squared at a given light quark
mass.  Since the parameters in our lattice simulations at different
lattice spacings are expressed in $r_1$ units, we require the physical
value of $r_1$ to convert to physical units and
take the physical pion mass and $\Delta^{(c)}$ from experiment.  Thus,
the $\approx 2.5\%$ uncertainty in $r_1^{\rm phys}$ gives a parametric
error in the chiral extrapolation.  Because the chiral extrapolation is
so mild, however, this error turns out to be negligible compared to other systematic errors.  Since we are taking the pion mass from experiment there is a negligible error due to the light quark mass uncertainty in the chiral extrapolation.  The strange sea quark mass enters the chiral extrapolation formulas, but the dependence is weak, and the error in the bare strange quark mass leads to a negligible parametric error in $h_{A_1}$.
The taste-splittings
$\Delta_\Xi$ have been determined in Ref.~\cite{Aubin:2004fs}, and their approximately
$10\%$ uncertainty also leads to a negligible error in $h_{A_1}(1)$.
The taste-violating hairpin coefficients have much larger fractional
uncertainties, but these too lead to a negligible uncertainty in
$h_{A_1}(1)$.  Even setting the rS$\chi$PT parameters to zero does not
change our result for $h_{A_1}(1)$ significantly. As mentioned above, our result does not change if we use
the continuum $\chi$PT formula in our chiral fits.

In the calculation of the form factor, the tadpole improved coefficient
$c_{SW}=1/u_0^3$ is obtained with $u_0$ from the Landau link on the
coarse lattices, but from the plaquette for $u_0$ on the fine and
coarser lattices.  Though unintentional, there is nothing wrong with
this, since it is not known {\it a priori} which provides the best
estimate of the tadpole improvement factor.  However, the $u_0$ term for
the spectator light (staggered) quark,
which appears in the tadpole improvement of the Asqtad action, was taken
from the Landau link on the coarse lattices, even though the sea quark
sector used $u_0$ from the plaquette.  On the fine and coarser lattices, $u_0$ was
taken to be the same in the light valence and sea quark sectors.  The
estimates of $u_0$ from plaquette versus Landau link differ only by
$4\%$ on the coarse lattices.

Although the effect of this mistuning is expected to be small (correcting $u_0$ would lead to a slightly different valence propagator and different tuned $\kappa$ values, thus leading to a small modification of the staggered chiral parameters in the
valence sector for the coarse lattices used as inputs to the chiral
fit), it is possible to study how much difference it makes using
the $h_{A_1}$ lattice data. 
Including all three lattice spacings and using our preferred chiral fit,
we find $h_{A_1}(1)=0.921(13)$ where the error here is
statistical only.  If we neglect the coarse data points, we find
$h_{A_1}(1)=0.920(17)$, almost unchanged except for a somewhat larger
statistical error.  We can also examine the ratios ${\cal R}_{\rm val}$
and ${\cal R}_{\rm sea}$.  In our preferred fit to all the lattice data these
are 0.9910(34) and 1.0059(90) respectively, where the errors are
again only statistical.  If we drop the coarse lattice data, these
become 0.9960(56) and 0.999(13) respectively.  Since the ratio ${\cal
R}_{\rm sea}$ has very little valence quark mass dependence, we can
combine ${\cal R}_{\rm sea}$ from the fit to all of the lattice data with ${\cal
R}_{\rm val}$ from the fit neglecting the coarse lattice data.  This is useful,
because ${\cal R}_{\rm sea}$ has the larger statistical error, so we
would like to use the full lattice data set to determine this ratio, thus
isolating the mistuning in the valence sector on the coarse lattices.
When this is done we find that the central value of the final
$h_{A_1}(1)$ is shifted upward by $0.4\%$, well within statistical
errors and smaller than our other systematic errors.  We assign a
systematic error of $0.4\%$ due to the $u_0$
mistuning.

\subsection{Finite volume effects}

The finite volume corrections to the integrals which appear in
heavy-light $\chi$PT formulas, including those for $B\to D^*$ were given
by Arndt and Lin~\cite{Arndt:2004bg}.  There are no new
integrals appearing in the staggered case, and it is straightforward to
use the results of Arndt and Lin in the rS$\chi$PT for $h_{A_1}(1)$, as
shown in Ref.~\cite{Laiho:2005ue}.  We find that although the finite
volume corrections in $h_{A_1}(1)$ would be large near the cusp at the
physical pion mass on the current MILC ensembles (ranging in size from
2.5-3.5 fm), for the less chiral data points at which we have actually
simulated, the finite volume effects are negligible.  For all data
points in our simulations the finite volume corrections are less than 1
part in $10^{4}$.  We therefore assign no error due to finite volume
effects.

\subsection{Discretization errors}

As shown in Ref.~\cite{ElKhadra:1996mp,Kronfeld:2000ck,Harada:2001fi,Harada:2001fj},
the matching of lattice gauge theory to QCD is accomplished by
normalizing the first few terms in the heavy-quark expansion.  This is
done by tuning the kinetic masses of the $D_s$ and $B_s$ mesons computed
using the SW action (for the heavy quarks) to the experimental meson
masses.  Tree-level tadpole-improved perturbation theory is used to tune the
coupling $c_{SW}$ and the rotation coefficient
$d_1$ for the bottom and charm quarks.  Once this matching is done, the
discretization errors in $h_{A_1}(1)$ are of order
$\alpha_s(\overline{\Lambda}/2m_Q)^2$ and $(\overline{\Lambda}/2m_Q)^3$
\cite{Kronfeld:2000ck}, where the powers of two are combinatoric factors.  The leading matching
uncertainty is of the order $\alpha_s(\overline{\Lambda}/2m_c)^2$.  We
estimate the size of this error setting $\alpha_s=0.3$,
$\overline{\Lambda}=500$ MeV, and $m_c=1.2$ GeV, which gives
$\alpha_s(\overline{\Lambda}/2m_c)^2=0.013$.

\begin{table}
\begin{center}
\caption{$h_{A_1}(1)$ at physical quark masses at different lattice spacings, where taste-violating effects have been removed, or shown to be negligible.  Discretization effects due to analytic terms
associated with the light quark sector and
heavy-quark discretization effects remain in the lattice data. }
 \label{tab:dis}
\begin{tabular}{cc}
\hline \hline
 $a$ (fm)\ \ & $h_{A_1}(1)$  \\
  \hline
  0.15  & 0.914(11)\\
   0.12 & 0.907(14) \\
   0.09 & 0.921(13)  \\
   \hline
\end{tabular}
\end{center}
\bigskip
\bigskip
\end{table}

\begin{figure}
\begin{center}
\includegraphics[scale=.55]{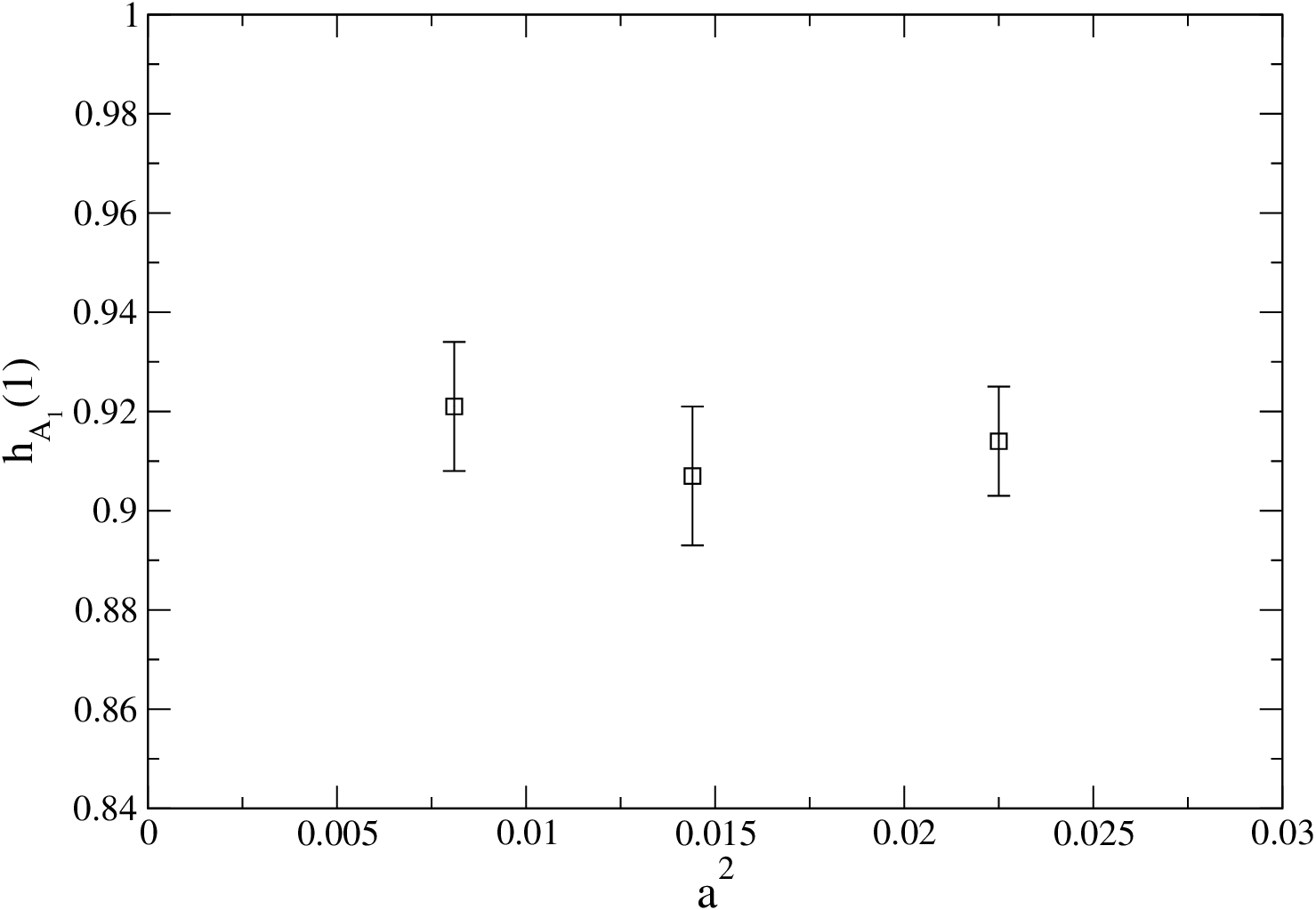}
\caption{$h_{A_1}(1)$ at physical quark masses versus $a^2$ (fm$^2$)
where taste-violating effects have been removed, or shown to be
negligible.  Discretization effects due to analytic terms associated
with the light quark sector and
heavy-quark discretization effects remain in the lattice data. \label{fig:aSq}}
\end{center}
\end{figure}

Since we have numerical data at three lattice spacings we are able to study how
well the power counting estimate accounts for observed discretization
effects.  Making use of Eq.~(\ref{eq:fid}), but varying the fiducial
lattice spacing from our lightest to coarsest lattices, we are able to
obtain $h_{A_1}(1)$ at physical quark masses, with discretization
effects associated with the staggered chiral logarithms removed in the
ratios appearing in Eq.~(\ref{eq:fid}).  The discretization effects that
remain are: taste-violations in $h_{A_1}^{\rm fid}$, taste violations at
higher order than NLO in the ratios, the effect of the analytic term coming from light
quark discretization effects (proportional to $\alpha_s a^2$), and the
heavy-quark discretization effects.  The taste-violations in
$h_{A_1}^{\rm fid}$ and the taste-violations in the ratios appearing at
higher order than NLO have been shown to be negligible.  We now consider
the remaining discretization errors coming from the light quark analytic
term and the heavy-quark discretization effects.  Table~\ref{tab:dis}
presents the results for $h_{A_1}(1)$ as obtained from
Eq.~(\ref{eq:fid}) and Figure~\ref{fig:aSq} shows them plotted as a
function of lattice spacing squared.  Although the Fermilab action 
and currents possess a smooth continuum limit, the MILC ensembles are not yet at small enough $a$ to obtain
simply $O(a)$ or $O(a^2)$ behavior.
The spread of the lattice data points gives some
indication of the size of the remaining discretization effects, however,
and we find that
the fine (0.09 fm) lattice data point and the coarse (0.12 fm) lattice data point differ
by $1.5\%$.  This is similar to our power counting estimate, and we
assign the larger of the two, $1.5\%$, as the systematic error due to
residual discretization effects.

\subsection{Summary}

Our final result, given the error budget in Table~\ref{tab:errors}, is 
\bea 
	h_{A_1}=0.921(13)(8)(8)(14)(6)(3)(4),
\eea
where the errors are statistical, parametric uncertainty in
$g_{DD^*\pi}$, chiral extrapolation errors, discretization errors,
parametric uncertainty in heavy-quark masses (kappa tuning),
perturbative matching, and the $u_0$ (mis)tuning on the coarse lattices.
Adding all systematic errors in quadrature, we obtain
\bea
	h_{A_1}(1) = 0.921(13)(20).
\eea
This final result differs slightly from that
presented at Lattice 2007
\cite{Laiho:2007pn}, where a preliminary $h_{A_1}(1)=0.924(12)(19)$ was
quoted.  There are three main changes in the analysis from the
preliminary result: our earlier result used a value of $\alpha_s$ in the
perturbative matching evaluated at the scale $2/a$, while the
present result uses the HLM \cite{Hornbostel:2002af} prescription to fix
the scale.  This causes a change of $0.1\%$, well within the estimated
systematic error due to the perturbative matching.  In the previous
result, the fine lattice data was blocked by 4 in the jackknife
procedure; we now block by 8 to fully account for autocorrelation
errors.  This does not change the central value, but increases the
statistical error slightly.  Finally, we have chosen a value for
$g_{DD^*\pi}=0.51\pm0.2$ instead of $g_{DD^*\pi}=0.45\pm0.15$ to be more
consistent with the range of values quoted in the literature.  This
causes a decrease in $h_{A_1}(1)$ of $0.2\%$.

\begin{table}
\begin{center}
\caption{Final error budget for $h_{A_1}(1)$ where each error is
discussed in the text.  Systematic errors are added in quadrature and
combined in quadrature with the statistical error to obtain the total
error.}
\label{tab:errors}
\begin{tabular}{cc}
\hline \hline
  Uncertainty & $h_{A_1}(1)$  \\
  \hline
  Statistics &  $1.4\%$   \\
  $g_{DD^*\pi}$ &  $0.9\%$  \\
  \ NLO vs NNLO $\chi$PT fits \ &  $0.9\%$  \\
  Discretization errors &  $1.5\%$  \\
  Kappa tuning &  $0.7\%$  \\
  Perturbation theory & $0.3\%$  \\
  $u_0$ tuning & $0.4\%$ \\
\hline 
   Total & $2.6\%$ \\
\hline \hline
\end{tabular}
\end{center}
\end{table}

\section{Conclusions}

We have introduced a new method to calculate the zero-recoil form factor
for the $B\to D^* \ell \nu$ decay.  We include 2+1 flavors of sea quarks
in the generation of the gauge ensembles, so the calculation is
completely unquenched.  We have introduced a new double ratio, which
gives the form factor directly, and leads to a large savings in the
computational cost.  The simulation is performed in a regime where we
expect rooted staggered chiral perturbation theory to apply; we
therefore use the rS$\chi$PT result for the $B\to D^*$ form factor
\cite{Laiho:2005ue} to perform the chiral extrapolation and to remove
taste-breaking effects.  To aid the chiral and continuum extrapolations,
we introduced a set of ratios that has allowed us to largely
disentangle light and heavy-quark discretization effects.  Our new
result, ${\cal F}(1)=h_{A_1}(1)=0.921(13)(20)$ is consistent with the
previous quenched result, ${\cal F}(1)=0.913^{+0.029}_{-0.034}$
\cite{Hashimoto:2001nb}, but our errors are both smaller and under
better theoretical control.  
This result allows us to
extract $|V_{cb}|$ from the experimental measurement of the $B\to D^*
\ell \nu$ form factor, which determines ${\cal F}(1)|V_{cb}|$.  After applying a $0.7\%$ electromagnetic correction to our value for ${\cal F}(1)$ \cite{Sirlin:1981ie}, and taking the most recent PDG average for $|V_{cb}|{\cal F}(1)=(35.9\pm 0.8)\times 10^{-3}$ \cite{Amsler:2008zz}, we find 
\bea |V_{cb}|=(38.7\pm 0.9_{\rm exp} \pm 1.0_{\rm theo})\times 10^{-3}.
\eea
This differs by about 2$\sigma$ from the inclusive determination $|V_{cb}|=(41.6\pm 0.6)\times 10^{-3}$ \cite{Amsler:2008zz}.  Our new value supersedes the previous Fermilab quenched number \cite{Hashimoto:2001nb}, as it should other quenched numbers such as that in Ref.~\cite{deDivitiis:2008df}\footnote{Ref.~\cite{deDivitiis:2008df} calculates the $B\to D^* \ell \nu$ form factor in the quenched approximation at zero and non-zero recoil momentum and uses a step-scaling method \cite{Guagnelli:2002jd} to control the heavy-quark discretization errors.}.
	
Our largest error in ${\cal F}(1)$ is the systematic error due to
heavy-quark discretization effects, which we have estimated using HQET
power counting and inspection of the numerical data at three lattice spacings.
This error can be reduced by going to finer lattice spacings, or by
using an improved Fermilab action \cite{Oktay:2008ex}.  When using this improved
action, it would be
necessary to improve the currents to the same order.  We have introduced a method for separating the heavy and light-quark discretization errors, where the physical $h_{A_1}$ can be factorized into two factors, $h_{A_1}^{\rm fid}\times{\cal R}_{\rm fid}$, such that the heavy quark discretization errors are largely isolated in $h_{A_1}^{\rm fid}$.  Combining our value of ${\cal R}_{\rm fid}=0.997(10)(13)$ (where the first error is statistical, and the second is due to systematics that do not cancel in the ratio) with a determination of $h_{A_1}^{\rm fid}$ at finer lattice spacings and/or with an improved action would be a cost-effective way of reducing the heavy-quark discretization errors. The next largest
error in our calculation of ${\cal F}(1)$ is statistical, and this error
drives many of the systematic errors.  This is mostly a matter of
computing.  It would also be desirable to perform the matching of the
heavy-quark current to higher order in perturbation theory, or by using
non-perturbative matching.  With these improvements, it would be
possible to bring the error in ${\cal F}(1)$ to or below $1\%$, allowing
a very precise determination of $|V_{cb}|$ from exclusive semi-leptonic
decays.

\acknowledgments

We thank Jon Bailey for a careful reading of the manuscript.
Computations for this work were carried out in part on facilities of
the USQCD Collaboration, which are funded by the Office of Science of
the U.S. Department of Energy; and on facilities of the NSF Teragrid under allocation TG-MCA93S002.
This work was supported in part by the United States Department of Energy
under Grant Nos.~DE--FC02-06ER41446 (C.D., L.L.), DE-FG02-91ER40661(S.G.), DE-FG02-91ER40677 (A.X.K.), DE-FG02-91ER40628  (C.B., J.L.), DE-FG02-04ER41298 (D.T.) and by the National Science Foundation under Grant Nos.~PHY-0555243, PHY-0757333, PHY-0703296 (C.D., L.L.), PHY-0555235 (J.L.), PHY-0456556 (R.S.).  R.T.E. and E.G. thank Fermilab and URA for their hospitality.
Fermilab is operated by Fermi Research Alliance, LLC, under Contract 
No.~DE-AC02-07CH11359 with the United States Department of Energy.

\appendix
\section{CHIRAL PERTURBATION THEORY}
\setcounter{equation}{0} \setcounter{section}{1}
\renewcommand{\theequation}{A\arabic{equation}}

Eq.~(34) of Ref.~\cite{Laiho:2005ue} gives the expression needed
for $h_{A_1}(1)$ in partially-quenched $\chi$PT with
degenerate up and down quark masses (the 2+1 case) in the rooted staggered theory:
\begin{eqnarray}\label{eq:schpt}  h_{A_1}^{(B_x)PQ,2+1}(1)/\eta_A&=& 1 + X_{A}(\Lambda_\chi)+\frac{g^2_{DD^*\pi}}{48
\pi^2f^2}\Bigg\{\frac{1}{16} \sum_{\begin{subarray}{l}j=xu,xu,xs\\ \Xi=I,P,4V,4A,6T \end{subarray}} \!\!\!\!\!\!\! \overline{F}_{j_\Xi} \nonumber\\
 &+& \frac{1}{3}\bigg[R^{[2,2]}_{X_I}\big(\{M^{(5)}_{X_I}\}
;\{\mu_I\}\big)\left(\frac{d\overline{F}_{X_I}}{dm^2_{X_I}}\right)
-\sum_{j \in
\{M^{(5)}_I\}}D^{[2,2]}_{j,X_I}\big(\{M^{(5)}_{X_I}\};\{\mu_I\}\big)\overline{F}_{j}
\bigg] \nonumber \\
  &+& a^2\delta'_{V}\bigg[R^{[3,2]}_{X_I}\big(\{M^{(7)}_{X_V}\}
;\{\mu_V\}\big)\left(\frac{d\overline{F}_{X_V}}{dm^2_{X_V}}\right)
-\sum_{j \in
\{M^{(7)}_V\}}D^{[3,2]}_{j,X_V}\big(\{M^{(7)}_{X_V}\};\{\mu_V\}\big)\overline{F}_{j}
 \Big] \nonumber \\
 &+& \big(V\rightarrow A\big)\Bigg\}, \end{eqnarray}
\noindent where
\begin{eqnarray}\label{eq:F}
F\left(m_j,z_j\right) &=& \frac{m_j^2}{z_j}\bigg\{z_j^3
\ln\frac{m_j^2}{\Lambda_\chi^2}+\frac{1}{3}z_j^3 -4z_j+2\pi \nonumber \\
&&
-\sqrt{z_j^2-1}(z_j^2+2)\left(\ln\Big[1-2z_j(z_j-\sqrt{z_j^2-1})\Big]-i\pi\right)\bigg\}
\nonumber
\\ && \longrightarrow (\Delta^{(c)})^2\ln\left(\frac{m_j^2}{\Lambda_\chi^2}\right)+{\cal
O}[(\Delta^{(c)})^3],
\end{eqnarray}
\noindent with $\overline{F}(m_j,z_j) = F(m_j,-z_j)$, and
$z_j=\Delta^{(c)}/m_j$, where $\Delta^{(c)}$ is the $D$-$D^*$ mass
splitting. The residues $R^{[n,k]}_j$ and $D^{[n,k]}_{j,i}$ are
defined in Refs.~\cite{Aubin:2003mg, Aubin:2003uc}, and for
completeness we quote them here:
\begin{eqnarray} R^{[n,k]}_j(\{M\},\{\mu\})  &\equiv & \frac{\prod_{a=1}^k
(\mu^2_a-m^2_j)}{\prod_{i\neq j} (m^2_i-m^2_j)}, \nonumber \\
D^{[n,k]}_{j, l}(\{M\},\{\mu\})  &\equiv &
-\frac{d}{dm^2_l}R^{[n,k]}_j(\{M\},\{\mu\}). \end{eqnarray}
\noindent These residues are a function of two sets of masses, the
numerator masses, $\{M\}=\{m_1, m_2,...,m_n \}$ and the
denominator masses, $\{\mu\}=\{ \mu_1, \mu_2,...,\mu_k \}$.  In
our 2+1 flavor case, we have
\begin{eqnarray} \{M_X^{(5)}\} &\equiv& \{m_\eta, m_X \}, \nonumber \\
    \{M_X^{(7)}\}&\equiv& \{m_\eta, m_{\eta'}, m_X \}, \nonumber \\
    \{\mu\}&\equiv&\{m_U,m_S\}. \end{eqnarray}
\noindent The masses $m_{\eta_I}$, $m_{\eta_V}$, $m_{\eta'_V}$ are
given by \cite{Aubin:2003mg}
\begin{eqnarray}
m^2_{\eta_I} & = & \frac{m^2_{U_I}}{3} + \frac{2m^2_{S_I}}{3},
\nonumber \\
 m^2_{\eta_V}  & = & \frac{1}{2}\left(m^2_{U_V}+m^2_{S_V}+\frac{3}{4}a^2\delta'_V - Z
    \right),\nonumber \\
    m^2_{\eta'_V}  & = & \frac{1}{2}\left(m^2_{U_V}+m^2_{S_V}+\frac{3}{4}a^2\delta'_V + Z
    \right),\nonumber \\
    Z & \equiv & \sqrt{(m^2_{S_V}-m^2_{U_V})^2 -
    \frac{a^2\delta'_V}{2}(m^2_{S_V}-m^2_{U_V})+\frac{9(a^2\delta'_V)^2}{16}}.
\end{eqnarray}
The ratio $R_{\rm val}^{\rm NLO}$ in the continuum through NLO in $\chi$PT is 
\begin{eqnarray}\label{eq:chptRval}  {\cal R}_{\rm val}^{\rm NLO}&=& 1 +\frac{g^2_{DD^*\pi}}{48
\pi^2f^2}\Bigg\{ \sum_{j=u,d,s}  \overline{F}_{xj} 
 + \frac{1}{3}\bigg[R^{[2,2]}_{X}\big(\{M^{(5)}_{X}\}
;\{\mu\}\big)\left(\frac{d\overline{F}_{X}}{dm^2_{X}}\right)
\nonumber\\ &-&\sum_{j \in
\{M^{(5)}_X\}}D^{[2,2]}_{j,X}\big(\{M^{(5)}_{X}\};\{\mu\}\big)\overline{F}_{j}
\bigg] 
- \sum_{j=u,d,s}  \overline{F}_{x'j} \nonumber\\
 &-& \frac{1}{3}\bigg[R^{[2,2]}_{X'}\big(\{M^{(5)}_{X'}\}
;\{\mu\}\big)\left(\frac{d\overline{F}_{X'}}{dm^2_{X'}}\right)
-\sum_{j \in
\{M^{(5)}_{X'} \}}D^{[2,2]}_{j,X'}\big(\{M^{(5)}_{X'}\};\{\mu\}\big)\overline{F}_{j}
\bigg] \Bigg\}, \end{eqnarray}
where 
\bea \{M_{X'}^{(5)}\} &\equiv& \{m_\eta, m_{X'} \},
\eea
and where $m_{X'}$ is a valence pion made of two quarks set to the fiducial valence quark mass, and the subscript $x'$ refers to a valence quark at the fiducial mass.  This ratio is one by construction when the valence quark mass equals the fiducial valence quark mass.



\bibliography{SuperBib}

\end{document}